\documentclass[1p]{elsarticle}

\usepackage{hyperref}
\usepackage{bm}
\usepackage{amsmath}
\usepackage{color}

\journal{Journal of Computational Physics}

\begin{document}

\begin{frontmatter}

\title{An Extension of Godunov SPH: Application to Negative Pressure Media}

\author[mymainaddress]{Keisuke Sugiura\corref{mycorrespondingauthor}}
\cortext[mycorrespondingauthor]{Corresponding author}
\ead{sugiura.keisuke@a.mbox.nagoya-u.ac.jp}

\author[mymainaddress]{Shu-ichiro Inutsuka}

\address[mymainaddress]{Department of Physics, Nagoya University, Aichi 464-8602, Japan}

\begin{abstract}
 The modification of Smoothed Particle Hydrodynamics (SPH) method with Riemann Solver is called Godunov SPH. We further extend the Godunov SPH to the description of a medium with negative pressure. Under certain circumstances, the SPH method shows an unphysical instability that results in particle clustering. This instability is called the tensile instability. The tensile instability occurs in positive pressure regions in a regular fluid if a very large number of neighbor particles are used with certain shapes of kernel functions, and it is significant in negative pressure regions that emerge in stretched elastic bodies. We must suppress the tensile instability in SPH for calculations of elastic bodies. In this study, we develop a new technique to remove the tensile instability by extending the Godunov SPH method and conducting a linear stability analysis of the equation of motion for the extended method. We find that the tensile instability can be suppressed by choosing an appropriate order of interpolation in the equation of motion of the Godunov SPH method. We also derive an analytic solution for a Riemann solver for a simple equation of state of an elastic body, and construct a Godunov SPH method for the equation of state that allows negative pressure.
\end{abstract}


\end{frontmatter}


\section{Introduction}
Smoothed Particle Hydrodynamics (SPH) is a computational fluid dynamics method. It is a Lagrangian method and does not require a Eulerian mesh. Each SPH particle mimics a fluid element and we can describe fluid dynamics by the motion of SPH particles. Lagrangian particle methods like SPH are suited for systems that have large voids or large deformed structures. Thus, the SPH method has been widely used in astrophysics and planetary science since it was proposed by Lucy \cite{Lucy1977} and Gingold $\&$ Monaghan \cite{Gingold-Monaghan1977}. The most popular form of SPH is called the standard SPH method.

Recently, some studies have attempted to apply SPH to elastic dynamics (e.g., \cite{Benz-Asphaug1994, Benz-Asphaug1995, Benz-Asphaug1999}). The application is straightforward because the equations of elastic dynamics are very similar to those of hydrodynamics. Furthermore, we can easily follow the crack history of elastic body elements because SPH is a Lagrangian particle method. Therefore SPH is a powerful tool for the calculation of disruptive collisions. 

However, the standard SPH method for elastic dynamics has one serious problem. As opposed to a compressed elastic body, which produces positive pressure, a stretched elastic body produces negative pressure. In the negative pressure regime, the standard SPH method has an instability that results in a clustering of particles. This instability is called the tensile instability, and was studied in detail by Swegle et al. \cite{Swegle-et-al1995}. It is also known that this tensile instability occurs even in the positive pressure regime. \cite{Dehnen-Aly2012} demonstrates that B-spline kernels produce the tensile instability, even in the positive pressure regime, if the number of neighbor particles is large. If the number of neighbor particles is small, we can dodge the tensile instability in the positive pressure, but this increases the `$E_{0}$ error' \cite{Read-Hayfield-and-Agertz2010}, which is the leading error of momentum equation, and will make erroneous result. Other discretizations such as \cite{Abel2011} successfully reduce this `$E_{0}$ error', but these discretization forms lose momentum conservation. These facts may suggest that the conservation and reordering property are not easily compatible with the method that does not suffer from the tensile instability. This instability is unphysical. Thus, for example, when we calculate collisions of asteroids without any prescription for the tensile instability, the size distribution of fragments may become completely wrong. Therefore, it is important to develop a method that suppresses the tensile instability.

There are many approaches toward the solution of the tensile instability in the standard SPH method (e.g., \cite{Randles-Libersky1996, Johnson-Beissel1996, Monaghan1999, Chen-et-al1999, Chen-et-al2015, Sigalotti-and-Lopez2008}). In \cite{Monaghan1999}, Monaghan introduced an artificial pressure that provides a strong repulsive force only when particles become close to each other. He also conducted a linear stability analysis for this method and found that this artificial pressure can suppress the instability at short wavelengths but does not affect the perturbations of long wavelengths. However, according to Mehra et al. \cite{Mehra2012}, this artificial pressure cannot suppress the tensile instability in simulations of hypervelocity impacts. 

Another formulation of SPH is called the Godunov SPH method \cite{Inutsuka2002}. It achieves the second-order spatial accuracy, whereas the standard SPH method does not have such a convergence property because of its rough approximation. The tensile instability does not occur if we solve the original equations of hydrodynamics or elastic dynamics exactly. Thus, this instability is caused by discretization error. We expect that if we use a method that has higher-order accuracy, such as the Godunov SPH method, we can suppress the tensile instability.

In this paper, we formulate a higher-order interpolation of the equations of the Godunov SPH method. Furthermore, we evaluate the stability of each order of interpolation by a linear stability analysis. A practical approach for dealing with the tensile instability using variable smoothing lengths is also shown. We also derive the analytical solution of the Riemann problem for a simple equation of state for an elastic body that allows negative pressure, and we construct a Godunov SPH method for this equation of state.

The structure of this paper is as follows: in Section 2, the essence of the SPH method and the equations for the Godunov SPH method are introduced. In Section 3, we present the motivation for higher-order interpolation, and then we derive the equations for higher-order interpolation. We also derive the analytical solution of the Riemann problem for a simple equation of state for an elastic body. In Section 4, we conduct a linear stability analysis of the equation of motion for the Godunov SPH method and evaluate its effectiveness for the tensile instability. Test calculations are presented in Section 5 and we show the validity of the analysis presented in Section 4. Section 6 is a summary of our work.

\section{SPH Method}

\subsection{Essence of SPH}
The contents of section 2.1 and 2.2 follow section 2 in \cite{Inutsuka2002}. The SPH method is a Lagrangian particle method, and hence we use the Lagrangian forms of the equation of motion and equation of energy for an inviscid fluid:

\begin{align}
&\frac{d\bm{v}}{dt}=-\frac{1}{\rho}\nabla P, \label{EoM-inviscid} \\ &\frac{du}{dt}=-\frac{P}{\rho}\nabla \cdot \bm{v}, \label{EoE-inviscid}
\end{align}

\noindent where $u$ is the specific internal energy, $\bm{v}$ is the velocity, $P$ is the pressure, and $\rho$ is the density.

In the SPH method, a physical quantity, $f$, at an arbitrary position is approximated by the convolution of nearby quantities. The convolution of a quantity $f$ at position $\bm{r}$ is defined as

\begin{equation}
\langle f \rangle(\bm{r})\equiv \int f(\bm{r}^{'})W(\bm{r}^{'}-\bm{r},h)d\bm{r}^{'},
\label{define-convolution}
\end{equation}

\noindent where $W(\bm{r}^{'}-\bm{r},h)$ is a kernel function and $h$ is a parameter called the smoothing length. We use the angle brackets $\langle \rangle$ to denote the convolution. We tentatively treat this smoothing length as constant. 

The kernel function has various forms; one of the simplest kernels is Gaussian,

\begin{equation}
W(\bm{r},h)=\Bigl[ \frac{1}{h\sqrt{\pi}} \Bigr]^{d} e^{-\bm{r}^{2}/h^{2}},
\label{kernel}
\end{equation}

\noindent where $d$ represents the spatial dimension. We use this Gaussian kernel throughout this paper.

Now we define the density at arbitrary position by the summation of the kernel function at particle positions,

\begin{equation}
\rho (\bm{r})\equiv \sum_{j}m_{j}W(\bm{r}-\bm{r}_{j},h).
\label{density}
\end{equation}

\noindent From this we can make the following identity:

\begin{equation}
1=\sum_{j}\frac{m_{j}}{\rho(\bm{r})}W(\bm{r}-\bm{r}_{j},h). 
\label{density-identity}
\end{equation}

Using the identity given by Eq.\,(\ref{density-identity}), and the definition of the convolution, Eq.\,(\ref{define-convolution}), we can express a physical quantity at particle position $i$ as

\begin{align}
f_{i}\equiv \langle f \rangle (\bm{r}_{i})&=\int f(\bm{r}^{'})W(\bm{r}^{'}-\bm{r}_{i},h)d\bm{r}^{'} \nonumber \\ &=\sum_{j} \int m_{j}\frac{f(\bm{r}^{'})}{\rho (\bm{r}^{'})}W(\bm{r}^{'}-\bm{r}_{i},h)W(\bm{r}^{'}-\bm{r}_{j},h)d\bm{r}^{'}.
\label{f-of-i}
\end{align}

\noindent Similarly, we can express the space derivative of a physical quantity of particle $i$ as

\begin{align}
\nabla f_{i} \equiv \Bigl< \frac{\partial f}{\partial \bm{r}} \Bigr>(\bm{r}_{i})&=\int f(\bm{r}^{'})\frac{\partial}{\partial \bm{r}_{i}}W(\bm{r}^{'}-\bm{r}_{i},h)d\bm{r}^{'} \nonumber \\ &=\sum_{j}\int m_{j}\frac{f(\bm{r}^{'})}{\rho (\bm{r}^{'})}\frac{\partial}{\partial \bm{r}_{i}}W(\bm{r}^{'}-\bm{r}_{i},h)W(\bm{r}^{'}-\bm{r}_{j},h)d\bm{r}^{'},  \label{dif-f-of-i}
\end{align}

\noindent where

\begin{equation}
\frac{\partial}{\partial \bm{r}_{i}}=\Bigl( \frac{\partial}{\partial x_{i}},\frac{\partial}{\partial y_{i}},\frac{\partial}{\partial z_{i}} \Bigr). 
\label{differ-i}
\end{equation}

\subsection{Standard SPH}

In the previous subsection we introduced the general formalism of the SPH method. However, the formalism that is used in the standard SPH method is simpler than Eq.\,(\ref{f-of-i}) and Eq.\,(\ref{dif-f-of-i}).

In the standard SPH method, we integrate Eq.\,(\ref{f-of-i}) using the approximation $W(\bm{r}^{'}-\bm{r}_{j},h) \approx \delta (\bm{r}^{'}-\bm{r}_{j})$, such that

\begin{equation}
f_{i} \approx \sum_{j}m_{j}\frac{f_{j}}{\rho_{j}}W(\bm{r}_{i}-\bm{r}_{j},h),
\label{f-of-i-standard}
\end{equation}

\noindent where $\delta (\bm{r})$ is the Dirac $\delta$ function. Similarly, we can integrate Eq.\,(\ref{dif-f-of-i}) and express the space derivative of a physical quantity of particle $i$ as

\begin{align}
\nabla f_{i} &= \sum_{j}\int m_{j}\frac{f(\bm{r}^{'})}{\rho (\bm{r}^{'})}\frac{\partial}{\partial \bm{r}_{i}}W(\bm{r}^{'}-\bm{r}_{i},h)W(\bm{r}^{'}-\bm{r}_{j},h)d\bm{r}^{'} \nonumber \\ &\approx \sum_{j}m_{j}\frac{f_{j}}{\rho_{j}}\frac{\partial}{\partial \bm{r}_{i}}W(\bm{r}_{i}-\bm{r}_{j},h). \label{dif-f-of-i-standard}
\end{align}

These expressions are derived from rough approximation, but in actual calculations Eq.\,(\ref{dif-f-of-i-standard}) is sufficient for the spatial derivative of a physical quantity.

\subsection{Godunov SPH}
In this subsection, we introduce the equation of motion and the equation of energy that were derived by \cite{Inutsuka2002}. We also introduce the equations for variable smoothing length in the Godunov SPH method.

\subsubsection{Equations for Godunov SPH}
The equation of motion for the Godunov SPH method is defined by the convolution of Eq.\,(\ref{EoM-inviscid}),

\begin{equation}
\dot{\bm{v}}_{i}\equiv \int \frac{d\bm{v}(\bm{r})}{dt}W(\bm{r}-\bm{r}_{i},h)d\bm{r}=-\int \frac{1}{\rho (\bm{r})}\nabla P(\bm{r})W(\bm{r}-\bm{r}_{i},h)d\bm{r},
\label{convolution-of-EoM}
\end{equation}

\noindent where the overdot shows a time derivative.

We transform the right-hand side of Eq.\,(\ref{convolution-of-EoM}) using Eq.\,(\ref{density-identity}) and integration by parts,

\begin{equation}
\dot{\bm{v}_{i}}=-\sum_{j}m_{j}\int \frac{P(\bm{r})}{\rho^{2}(\bm{r})}\Bigl[ \frac{\partial}{\partial \bm{r}_{i}}-\frac{\partial}{\partial \bm{r}_{j}} \Bigr]W(\bm{r}-\bm{r}_{i},h)W(\bm{r}-\bm{r}_{j},h)d\bm{r}.
\label{EoM-of-Godunov}
\end{equation}

If we multiply both sides of Eq.\,(\ref{EoM-of-Godunov}) by the mass of particle $i$, the left-hand side becomes the time derivative of the momentum of particle $i$, and the right-hand side becomes antisymmetric with respect to $i$ and $j$. Therefore, in this formalism, as in standard SPH, the linear momentum and angular momentum of a particle system are conserved.

When the equation of state only depends on density, or the fluid is barotropic, we can follow the evolution of the fluid using only the equation of motion. However, in the case of a general equation of state like $P=P(\rho ,u)$, or a fluid that has  a shock wave, the equation of energy is also required. The equation of energy for the Godunov SPH method is derived from the convolution of Eq.\,(\ref{EoE-inviscid}), similarly to the derivation of Eq.\,(\ref{convolution-of-EoM}),

\begin{equation}
\dot{u}_{i} \equiv \int \frac{du(\bm{r})}{dt}W(\bm{r}-\bm{r}_{i},h)d\bm{r} = -\int \frac{P(\bm{r})}{\rho (\bm{r})}[\nabla \cdot \bm{v}(\bm{r})]W(\bm{r}-\bm{r}_{i},h)d\bm{r}.
\label{convolution-of-EoE}
\end{equation}

\noindent We integrate the right-hand side of Eq.\,(\ref{convolution-of-EoE}) by parts and find

\begin{align}
&-\int \frac{P(\bm{r})}{\rho (\bm{r})}[\nabla \cdot \bm{v}(\bm{r})]W(\bm{r}-\bm{r}_{i},h)d\bm{r} \nonumber \\ &= -\int \frac{1}{\rho (\bm{r})}[\nabla \cdot P\bm{v}]W(\bm{r}-\bm{r}_{i},h)d\bm{r} +\int \frac{1}{\rho (\bm{r})}[\bm{v} \cdot \nabla P]W(\bm{r}-\bm{r}_{i},h)d\bm{r}. \label{convolution-of-EoE-2}
\end{align}

\noindent Then, we use the following approximation:

\begin{equation}
\int \frac{1}{\rho (\bm{r})}[\bm{v}\cdot \nabla P]W(\bm{r}-\bm{r}_{i},h)d\bm{r}\approx \int \frac{1}{\rho (\bm{r})}[\dot{\bm{r}}_{i}\cdot \nabla P]W(\bm{r}-\bm{r}_{i},h)d\bm{r}.
\label{convolution-of-EoE-3}
\end{equation}

\noindent By using Eqs.\,(\ref{convolution-of-EoE-3}) and (\ref{density-identity}) we can transform Eq.\,(\ref{convolution-of-EoE-2}) into

\begin{equation}
\dot{u}_{i}=-\sum_{j}m_{j}\int \frac{P(\bm{r})}{\rho^{2}(\bm{r})}[\bm{v}(\bm{r})-\dot{\bm{r}}_{i}]\cdot \Bigl[ \frac{\partial}{\partial \bm{r}_{i}}-\frac{\partial}{\partial \bm{r}_{j}} \Bigr] W(\bm{r}-\bm{r}_{i},h)W(\bm{r}-\bm{r}_{j},h)d\bm{r}.
\label{EoE-of-Godunov}
\end{equation}

Equations (\ref{EoM-of-Godunov}) and (\ref{EoE-of-Godunov}) are not yet useful for practical calculation because they contain spatial integration. Thus further approximations are required.

\subsubsection{Convolution}
The equation of motion and the equation of energy for the Godunov SPH method are shown in Eq.\,(\ref{EoM-of-Godunov}) and Eq.\,(\ref{EoE-of-Godunov}), and these equations involve spatial integration. The $\rho (\bm{r})$ parts of the integrands are given by Eq.\,(\ref{density}). Thus, we cannot integrate Eq.\,(\ref{EoM-of-Godunov}) and Eq.\,(\ref{EoE-of-Godunov}) analytically. Furthermore, it is almost impossible to integrate these equations numerically for each pair of $i$-th and $j$-th particles in practical calculations because of the heavy numerical cost. Therefore, we must interpolate and assume the spatial distribution of physical quantities like $\rho (\bm{r})$.

We now define a new convenient coordinate system for the integration. This coordinate system has its origin at $(\bm{r}_{i}+\bm{r}_{j})/2$, and we define the $s$-axis to be along the direction of the vector $\bm{r}_{i}-\bm{r}_{j}$. We use $\bm{s}_{\perp}$ to denote the component of vector $\bm{r}$ that is perpendicular to the $s$-axis, and we define $\bm{e}_{ij}\equiv (\bm{r}_{i}-\bm{r}_{j})/|\bm{r}_{i}-\bm{r}_{j}|$ as the unit vector along the $s$-axis, and $\Delta s_{ij}\equiv |\bm{r}_{i}-\bm{r}_{j}|$ as the distance between particles $i$ and $j$.

Essentially, Eq.\,(\ref{EoM-of-Godunov}) and Eq.\,(\ref{EoE-of-Godunov}) include the integration as follows:

\begin{equation}
\int \frac{f(\bm{r})}{\rho^{2} (\bm{r})}W(\bm{r}-\bm{r}_{i},h)W(\bm{r}-\bm{r}_{j},h)d\bm{r}.
\label{general-convolution}
\end{equation}

\noindent For simplicity we define the weighted average $f_{ij}^{\ast}$ as

\begin{equation}
\int \frac{f(\bm{r})}{\rho^{2} (\bm{r})}W(\bm{r}-\bm{r}_{i},h)W(\bm{r}-\bm{r}_{j},h)d\bm{r}=f_{ij}^{\ast}\int \frac{1}{\rho^{2}(\bm{r})}W(\bm{r}-\bm{r}_{i},h)W(\bm{r}-\bm{r}_{j},h)d\bm{r}.
\label{weighted-average-of-f}
\end{equation}

\noindent Then we expand $\rho^{-2}(\bm{r})$ linearly in the direction perpendicular to the $s$-axis as

\begin{equation}
\rho^{-2}(\bm{r})\approx \rho^{-2}(s)+\bm{s}_{\perp}\cdot \nabla \rho^{-2}(s).
\label{perpendicular-rho2}
\end{equation}

\noindent Note that if we use the approximation of Eq.\,(\ref{perpendicular-rho2}) the component that is perpendicular to the $s$-axis vanishes because of the symmetric property of the kernel function. By using Eq.\,(\ref{perpendicular-rho2}) we can transform Eq.\,(\ref{weighted-average-of-f}) into

\begin{equation}
\int \frac{f(\bm{r})}{\rho^{2}(\bm{r})}W(\bm{r}-\bm{r}_{i},h)W(\bm{r}-\bm{r}_{j},h)d\bm{r}=f_{ij}^{\ast}V_{ij}^{2}W(\bm{r}_{i}-\bm{r}_{j},\sqrt{2}h),
\label{convolution-final}
\end{equation}

\noindent where

\begin{equation}
V_{ij}^{2}=\int_{-\infty}^{\infty} \frac{\sqrt{2}}{h\sqrt{\pi}}\frac{1}{\rho^{2}(s)}\exp \Bigl(-\frac{2s^{2}}{h^{2}} \Bigr) ds.
\label{define-Vij2}
\end{equation}

Finally, if we interpolate $\rho(s)$ in the direction of the $s$-axis, we can integrate $V_{ij}^{2}$ analytically. Here it is convenient to define the specific volume and its gradient as

\begin{align}
&V(\bm{r})=\frac{1}{\rho (\bm{r})}, \nonumber \\ &\nabla V(\bm{r})=-\frac{1}{\rho^{2}(\bm{r})}\nabla \rho (\bm{r})=-\frac{1}{\rho^{2}(\bm{r})}\sum_{j}m_{j}\nabla W(\bm{r}-\bm{r}_{j},h). \label{specific-volume}
\end{align}

The simplest interpolation of $V(s)$ is linear interpolation, which we express as

\begin{equation}
V(s)=\frac{1}{\rho (s)}=C_{ij}s+D_{ij},
\label{linear-interpolation}
\end{equation}

\noindent where

\begin{align}
&C_{ij}=\frac{V(\bm{r}_{i})-V(\bm{r}_{j})}{\Delta s_{ij}}, \nonumber \\ &D_{ij}=\frac{V(\bm{r}_{i})+V(\bm{r}_{j})}{2}. \label{linear-coefficients}
\end{align}

\noindent By substituting Eq.\,(\ref{linear-interpolation}) into Eq.\,(\ref{define-Vij2}), we obtain

\begin{equation}
V_{ij,{\rm linear}}^{2}(h)=\frac{1}{4}h^{2}C_{ij}^{2}+D_{ij}^{2}.
\label{Vij2-linear}
\end{equation}

Another convenient interpolation is cubic spline interpolation. This is expressed as

\begin{equation}
V(s)=\frac{1}{\rho (s)}=A_{ij}s^{3}+B_{ij}s^{2}+C_{ij}s+D_{ij},
\label{cubic-spline-interpolation}
\end{equation}

\noindent where

\begin{align}
&A_{ij}=-2\frac{V_{i}-V_{j}}{\Delta s_{ij}^{3}}+\frac{V_{i}^{'}+V_{j}^{'}}{\Delta s_{ij}^{2}}, \nonumber \\ &B_{ij}=\frac{1}{2}\frac{V_{i}^{'}-V_{j}^{'}}{\Delta s_{ij}},\nonumber \\ &C_{ij}=\frac{3}{2}\frac{V_{i}-V_{j}}{\Delta s_{ij}}-\frac{1}{4}(V_{i}^{'}+V_{j}^{'}), \nonumber \\ &D_{ij}=\frac{1}{2}(V_{i}+V_{j})-\frac{1}{8}(V_{i}^{'}-V_{j}^{'})\Delta s_{ij}, \nonumber \\ &V_{i}=V(\bm{r}_{i}), \nonumber \\ &V_{j}=V(\bm{r}_{j}), \nonumber \\ &V_{i}^{'}=\bm{e}_{ij}\cdot \nabla V(\bm{r}_{i}), \nonumber \\ &V_{j}^{'}=\bm{e}_{ij}\cdot \nabla V(\bm{r}_{j}). \label{cubic-coefficients}
\end{align}

\noindent Then we can calculate Eq.\,(\ref{define-Vij2}) and obtain

\begin{equation}
V_{ij,{\rm cubic}}^{2}(h)=\frac{15}{64}h^{6}A_{ij}^{2}+\frac{3}{16}h^{4}(2A_{ij}C_{ij}+B_{ij}^{2})+\frac{1}{4}h^{2}(2B_{ij}D_{ij}+C_{ij}^{2})+D_{ij}^{2}.
\label{Vij2-cubic}
\end{equation}

If we substitute Eq.\,(\ref{Vij2-linear}) or Eq.\,(\ref{Vij2-cubic}) and Eq.\,(\ref{convolution-final}), into Eq.\,(\ref{EoM-of-Godunov}) and Eq.\,(\ref{EoE-of-Godunov}), the equation of motion and the equation of energy finally become

\begin{align}
&\dot{\bm{v}}_{i}=-2\sum_{j}m_{j}P_{ij}^{\ast}V_{ij}^{2}(h)\frac{\partial}{\partial \bm{r}_{i}}W(\bm{r}_{i}-\bm{r}_{j},\sqrt{2}h), \label{EoM-consth} \\ &\dot{u}_{i}=-2\sum_{j}m_{j}(P_{ij}^{\ast}\bm{v}_{ij}^{\ast}-P_{ij}^{\ast}\dot{\bm{r}}_{i})\cdot V_{ij}^{2}(h)\frac{\partial}{\partial \bm{r}_{i}}W(\bm{r}_{i}-\bm{r}_{j},\sqrt{2}h). \label{EoE-consth}
\end{align} 

In \cite{Inutsuka2002}, Inutsuka introduced a Riemann solver that uses the physical quantities of the $i$-th and $j$-th particles as initial conditions of the shock tube problem. We use this Riemann solver for $P_{ij}^{\ast}$ and $\bm{v}_{ij}^{\ast}$ and can introduce necessary but minimal viscosity to deal with shock waves. However, in this paper, in order to separate the effect of viscosity and the effectiveness of the formalism of the Godunov SPH method against the tensile instability, we use Eq.\,(\ref{P-inviscid}) for $P_{ij}^{\ast}$ when we analyze the stability,

\begin{equation}
P_{ij}^{\ast}=\frac{P_{i}+P_{j}}{2}.
\label{P-inviscid}
\end{equation}

\noindent If we use Eq.\,(\ref{P-inviscid}) instead of the Riemann solver, we can turn off the viscosity. 

Throughout this paper we use the following simple equation of state (e.g., \cite{Monaghan1999}):

\begin{equation}
P=C_{s}^{2}(\rho - \rho_{0,{\rm eos}}),
\label{EoS}
\end{equation}

\noindent Thus, we do not use the equation of energy given by Eq.\,(\ref{EoE-consth}). Here, $\rho_{0,{\rm eos}}$ is a reference density and $C_{s}$ is the sound velocity, which we treat as a constant. 

\subsubsection{Variable smoothing length}
We have so far treated the smoothing length as constant. The smoothing length should be close to the average particle spacing, and the average particle spacing varies largely in space when the density varies largely. In practical calculations we should vary the smoothing length according to the local mean of the particle spacing. In this section, the smoothing length is represented by spatial variable $h(\bm{r})$. Then the integration of Eq.\,(\ref{EoM-of-Godunov}) and Eq.\,(\ref{EoE-of-Godunov}) includes $h(\bm{r})$, and we cannot integrate analytically even if we use the polynomial approximation of $\rho^{-2}(\bm{r})$.

In \cite{Inutsuka2002}, Inutsuka used approximate analytical integration assuming that the smoothing length is $h_{i}$ for the half of the integration space that includes particle $i$, and $h_{j}$ for the other half. Then the equation of motion and the equation of energy for variable smoothing length become

\begin{align}
\dot{\bm{v}}_{i}=&-\sum_{j}m_{j}P_{ij}^{\ast}\Bigl[ V_{ij}^{2}(h_{i})\frac{\partial}{\partial \bm{r}_{i}}W(\bm{r}_{i}-\bm{r}_{j},\sqrt{2}h_{i})\nonumber \\ &+V_{ij}^{2}(h_{j})\frac{\partial}{\partial \bm{r}_{i}}W(\bm{r}_{i}-\bm{r}_{j},\sqrt{2}h_{j})\Bigr], \label{EoM-variableh} \\ \dot{u}_{i}=&-\sum_{j}m_{j}(P_{ij}^{\ast}\bm{v}_{ij}^{\ast}-P_{ij}^{\ast}\dot{\bm{r}}_{i})\cdot \Bigl[ V_{ij}^{2}(h_{i})\frac{\partial}{\partial \bm{r}_{i}}W(\bm{r}_{i}-\bm{r}_{j},\sqrt{2}h_{i})\nonumber \\ &+V_{ij}^{2}(h_{j})\frac{\partial}{\partial \bm{r}_{i}}W(\bm{r}_{i}-\bm{r}_{j},\sqrt{2}h_{j})\Bigr]. \label{EoE-variableh}
\end{align}

\noindent This approximation assumes that the smoothing length does not vary much within the neighborhood of particles $i$ and $j$. 

A possible way to decide the smoothing length of particle $i$ is, for example,

\begin{equation}
h_{i}=\eta \Bigl[ \frac{m_{i}}{\rho^{\ast}_{i}} \Bigr]^{1/d},
\label{variable-h}
\end{equation}

\noindent where

\begin{equation}
\rho_{i}^{\ast}=\sum_{j}m_{j}W(\bm{r}_{i}-\bm{r}_{j},h_{i}^{\ast}), \ \ h_{i}^{\ast}=h_{i}C_{{\rm smooth}},
\label{Csmooth}
\end{equation}

\noindent and $C_{{\rm smooth}}$ is a constant to smooth the spatial variation of the smoothing length, $\eta$ is a constant to determine the ratio of the smoothing length to the average particle spacing. In this study we use $\eta =1$. $C_{{\rm smooth}}>1$ we can obtain smoother variation of the smoothing length than that of the density, and the approximation that was used to derive Eq.\,(\ref{EoM-variableh}) and Eq.\,(\ref{EoE-variableh}) becomes more accurate. However, the number of neighbor particles to obtain correct smoothing length becomes large if $C_{{\rm smooth}}$ becomes large. Eq.\,(\ref{variable-h}) and Eq.\,(\ref{Csmooth}) are recursive equations that require iterative calculation of $h_{i}$. In practical calculations the use of $h_{i}$ in the previous time step for $h_{i}$ in Eq.\,(\ref{Csmooth}) works well.

\section{Further Extension of Godunov SPH}
In this section, we show that the original formalism of the Godunov SPH method is in principle stable against the tensile instability. Next, we derive an equation for higher-order interpolation of $V_{ij}^{2}$ for use in Eq.\,(\ref{EoM-of-Godunov}). We also derive an analytical solution of the Riemann solver for Eq.\,(\ref{EoS}) so that we can use the Godunov SPH method for an equation of state that allows negative pressure.

\subsection{Motivation}
The equation of motion for the Godunov SPH method is given by Eq.\,(\ref{EoM-of-Godunov}) before we interpolate the distribution of physical quantities. To derive this equation we only take the convolution of the original equation of motion for an inviscid fluid. This convolution only introduces smoothing of the physical quantity, and thus should not cause instability. This suggests that the exact integration of Eq.\,(\ref{EoM-of-Godunov}) should remove the tensile instability even in the case of negative pressure. Note that the density at arbitrary position is given by Eq.\,(\ref{density}) and the pressure at arbitrary position is calculated from Eq.\,(\ref{EoS}). Thus we can calculate the convolution integral in Eqs.\,(\ref{EoM-of-Godunov}) and (\ref{define-Vij2}) using numerical integration.

To show this, we derive the dispersion relation for Eq.\,(\ref{EoM-of-Godunov}). For simplicity, we focus on the one-dimensional case and calculate the dispersion relation almost exactly using the method shown in Appendix A. In short, we assign particles with equal spacing and add a sinusoidal perturbation with wave number $k$ and calculate the acceleration. Then we can calculate the squared angular frequency $\omega^{2}$ by taking the ratio of the displacement and acceleration.

\begin{figure}[!htb]
 \begin{center}
 \includegraphics[width=9cm,height=6cm]{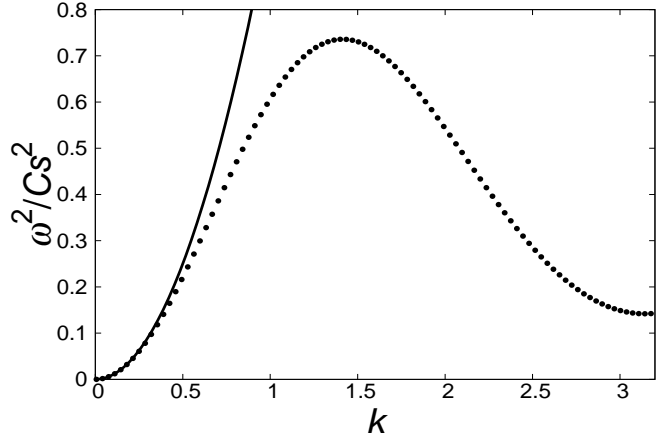}
 \caption{Dispersion relation of Eq.\,(\ref{EoM-of-Godunov}) for negative pressure. The horizontal axis is $k$, the vertical axis is $\omega^{2} /C_{s}^{2}$, and the unit of length is normalized by the average particle spacing. Solid curve shows the exact dispersion relation of the sound wave.}
 \label{DR-numerical-integration}
 \end{center}
\end{figure}

Figure \ref{DR-numerical-integration} shows the dispersion relation of Eq.\,(\ref{EoM-of-Godunov}) that is calculated using the method of Appendix A. Here we consider all particles to have equal mass, and the parameters are as follows: mass of each particle $m=0.0008$, particle spacing in the unperturbed state $\Delta x=0.001$, $h=0.001$, $C_{s}=1.0$, $\rho_{0,{\rm eos}}=1.0$. The average density is $\rho_{0}=0.8$ and the average pressure is $P_{0}=-0.2$ (negative). Numerical integration is done using a simple trapezoidal law, and the integral interval is $\Delta x/10$. The unit of length is normalized by the average particle spacing. Thus, $k=\pi$ corresponds to a perturbation wavelength of two particle spacings. Negative $\omega^{2}$ means the perturbation of the given wave number is unstable.

As shown in Fig.\,\ref{DR-numerical-integration}, $\omega^{2}>0$ for all wave numbers. Thus, we can see that the exact integration of Eq.\,(\ref{EoM-of-Godunov}) can remove the tensile instability for negative pressure. 

\begin{figure}[!htb]
 \begin{center}
 \includegraphics[width=9cm,height=6cm]{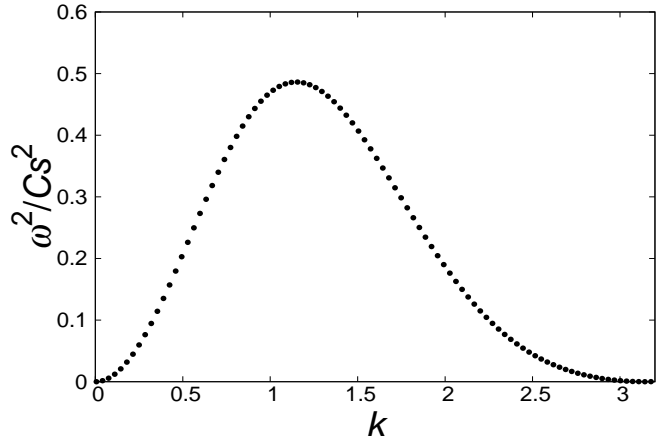}
 \caption{Dispersion relation of Eq.\,(\ref{EoM-consth}) for negative pressure, where Eq.\,(\ref{define-Vij2}) is integrated numerically.}
 \label{DR-numerical-Vij2}
 \end{center}
\end{figure}

Next, we examine a simpler case. In this case we use the equation of motion given by Eq.\,(\ref{EoM-consth}), but we integrate Eq.\,(\ref{define-Vij2}) numerically. Here we use $P_{ij}^{\ast}$ given by Eq.\,(\ref{P-inviscid}). This is the same as when we integrate Eq.\,(\ref{EoM-of-Godunov}), but the pressure is assumed constant, $P(\bm{r})=(P_{i}+P_{j})/2$. Figure \ref{DR-numerical-Vij2} shows the dispersion relation when we calculate the acceleration using this method, where all parameters are the same as in the case of Fig.\,\ref{DR-numerical-integration}.

In Fig.\,\ref{DR-numerical-Vij2} it appears that $\omega^{2} =0$ at around $k=\pi$, but actually these $\omega^{2}$ are small value and not 0. Thus, our calculations are free from the instability, even for negative pressure, if we integrate $V_{ij}^{2}$ numerically.

As shown in the test calculations in section 5, the Godunov SPH method also suffers from the tensile instability in the case of inappropriate interpolation methods. Therefore, the exact integration of the convolution can suppress the tensile instability. These results imply that the tensile instability of the Godunov SPH method comes from the errors in the approximation for integration. Thus we expect that if we improve the interpolation of $V_{ij}^{2}$ we can remove the tensile instability.

\subsection{Quintic spline interpolation}
As we discussed, if we use a higher-order interpolation of $V_{ij}^{2}$ to achieve $V_{ij}^{2}$ closer to the exact integration of Eq.\,(\ref{define-Vij2}), it is possible to suppress the tensile instability. In this subsection, we derive an equation for the higher-order interpolation.

In \cite{Inutsuka2002}, Inutsuka constructed a cubic spline interpolation of $V(s)$ by using four quantities: the specific volume and its derivative for particles $i$ and $j$. We can construct a quintic spline interpolation by adding two more quantities: the second derivative of the specific volume for particles $i$ and $j$.

To derive the second derivative of the specific volume for an arbitrary direction, we simply use Eq.\,(\ref{dif-f-of-i-standard}) and differentiate the first derivative to obtain

\begin{equation}
\frac{\partial^{2} V(\bm{r}_{i})}{\partial x_{\alpha}\partial x_{\beta}} = \sum_{j}\frac{m_{j}}{\rho_{j}}\frac{\partial V(\bm{r}_{j})}{\partial x_{\alpha,j}}\frac{\partial}{\partial x_{\beta,i}}W(\bm{r}_{i}-\bm{r}_{j},h),
\label{second-differencial-V}
\end{equation}

\noindent where $\alpha$ and $\beta$ take values 1, 2, or 3, and $x_{1}=x$, $x_{2}=y$, $x_{3}=z$. Actually, the second derivative that is calculated with Eq.\,(\ref{second-differencial-V}) is not so accurate and we obtain a smoother second derivative than the exact value, but this method is sufficient for the tensile instability as we will show using the linear stability analysis of Section 4. In the actual simulations, we calculate the second derivative using Eq.\,(\ref{second-differencial-V}) after calculating the first derivative using Eq.\,(\ref{specific-volume}).

Once we calculate the second derivative, we can construct the quintic spline interpolation of $V(s)$. This is expressed as

\begin{equation}
V(s)=A_{ij}s^{5}+B_{ij}s^{4}+C_{ij}s^{3}+D_{ij}s^{2}+E_{ij}s+F_{ij},
\label{quintic-spline-interpolation}
\end{equation}

\noindent where

\begin{align}
&A_{ij}=6\frac{V_{i}-V_{j}}{\Delta s_{ij}^{5}}-3\frac{V_{i}^{'}+V_{j}^{'}}{\Delta s_{ij}^{4}}+\frac{1}{2}\frac{V_{i}^{''}-V_{j}^{''}}{\Delta s_{ij}^{3}}, \nonumber \\ & B_{ij}=-\frac{1}{2}\frac{V_{i}^{'}-V_{j}^{'}}{\Delta s_{ij}^{3}}+\frac{1}{4}\frac{V_{i}^{''}+V_{j}^{''}}{\Delta s_{ij}^{2}}, \nonumber \\ & C_{ij}=-5\frac{V_{i}-V_{j}}{\Delta s_{ij}^{3}}+\frac{5}{2}\frac{V_{i}^{'}+V_{j}^{'}}{\Delta s_{ij}^{2}}-\frac{1}{4}\frac{V_{i}^{''}-V_{j}^{''}}{\Delta s_{ij}}, \nonumber \\ &D_{ij}=\frac{3}{4}\frac{V_{i}^{'}-V_{j}^{'}}{\Delta s_{ij}}-\frac{1}{8}(V_{i}^{''}+V_{j}^{''}), \nonumber \\ &E_{ij}=\frac{15}{8}\frac{V_{i}-V_{j}}{\Delta s_{ij}}-\frac{7}{16}(V_{i}^{'}+V_{j}^{'})+\frac{1}{32}(V_{i}^{''}-V_{j}^{''})\Delta s_{ij}, \nonumber \\ &F_{ij}=\frac{1}{2}(V_{i}+V_{j})-\frac{5}{32}(V_{i}^{'}-V_{j}^{'})\Delta s_{ij}+\frac{1}{64}(V_{i}^{''}+V_{j}^{''})\Delta s_{ij}^{2}, \label{quintic-coefficients}
\end{align}

\noindent and

\begin{align}
V^{''}_{i}=\sum_{\alpha}\sum_{\beta}e_{ij,\alpha}e_{ij,\beta}\frac{\partial^{2}V(\bm{r}_{i})}{\partial x_{\alpha}\partial x_{\beta}}, \nonumber \\ V^{''}_{j}=\sum_{\alpha}\sum_{\beta}e_{ij,\alpha}e_{ij,\beta}\frac{\partial^{2}V(\bm{r}_{j})}{\partial x_{\alpha}\partial x_{\beta}}. \label{quintic-coefficients-2}
\end{align}

\noindent Here, $e_{ij,\alpha}$ shows the $\alpha$ direction component of $\bm{e}_{ij}$. By substituting Eq.\,(\ref{quintic-spline-interpolation}) into Eq.\,(\ref{define-Vij2}) we find

\begin{align}
V_{ij,{\rm quintic}}^{2}(h)&=\frac{945}{1024}h^{10}A_{ij}^{2}+\frac{105}{256}h^{8}(2A_{ij}C_{ij}+B_{ij}^{2})+\frac{15}{64}h^{6}(2A_{ij}E_{ij}+2B_{ij}D_{ij}+C_{ij}^{2}) \nonumber \\ & +\frac{3}{16}h^{4}(2B_{ij}F_{ij}+2C_{ij}E_{ij}+D_{ij}^{2})+\frac{1}{4}h^{2}(2D_{ij}F_{ij}+E_{ij}^{2})+F_{ij}^{2}. \label{Vij2-quintic}
\end{align}

\subsection{Riemann solver for elastic equation of state}
Realistic equations of state generate negative pressure under certain circumstances. To develop a Godunov SPH method for such an equation of state we should derive the appropriate solution of a Riemann problem that includes shock waves with negative pressure. The emergence of a shock wave with negative pressure may not be obvious. As explained in many textbooks (e.g., \cite{Landau-Lifshitz1987}), a shock wave is generated by nonlinear steepening of a sound wave of finite amplitude. This nonlinear effect exists even in negative pressure media, and we have confirmed steepening in a negative pressure state for Eq.\,(\ref{EoS}) (Fig.\,\ref{steepening-of-wave} of Section 5). This fact implies the emergence of a shock wave with negative pressure. Therefore, it is valuable to construct a Riemann solver for such an equation of state and develop a Godunov SPH method that can handle shock waves in negative pressure regions.

In this subsection, we derive the analytical solution of a Riemann problem that uses the equation of state of Eq.\,(\ref{EoS}). A method to solve a Riemann problem is called a Riemann solver. The Riemann solver for an ideal gas has been extensively analyzed in \cite{Leer1978}. Godunov SPH uses the resultant pressure $P^{\ast}$ and velocity $\bm{v}^{\ast}$ of the Riemann problem for $P_{ij}^{\ast}$ and $\bm{v}_{ij}^{\ast}$ of the equation of motion, Eq.\,(\ref{EoM-consth}), and the equation of energy, Eq.\,(\ref{EoE-consth}). 

Let us consider a shock wave propagating with velocity $v_{s}$ in a medium of density $\rho$ moving with velocity $v$. According to the Rankine-Hugoniot relations, we have the following relations across the shock wave:

\begin{align}
&\pm W_{s}(V^{\ast}-V)+(v^{\ast}-v)=0, \label{Rankine-Hugoniot-mass} \\ &\pm W_{s}(v^{\ast}-v)-(P^{\ast}-P)=0, \label{Rankine-Hugoniot-momentum}
\end{align}

\noindent where $W_{s}$ denotes the Lagrangian shock speed $|\rho v_{s}|$, the sign shows the direction of shock propagation, and the quantities marked by asterisks correspond to the post-shock values. Eq.\,(\ref{Rankine-Hugoniot-mass}) and Eq.\,(\ref{Rankine-Hugoniot-momentum}) represent conservation of mass and momentum flux across the shock wave. Using Eq.\,(\ref{Rankine-Hugoniot-mass}), Eq.\,(\ref{Rankine-Hugoniot-momentum}), and the equation of state for the post-shock value,

\begin{equation}
P^{\ast}=C_{s}^{2}(\rho^{\ast}-\rho_{0,{\rm eos}}), 
\label{EoS-postshock}
\end{equation}

\noindent we can express $W_{s}$ as

\begin{equation}
W_{s}=C_{s}\sqrt{\rho \Bigl(\rho + \frac{P^{\ast}-P}{C_{s}^{2}}\Bigr)}. 
\label{W-shock}
\end{equation}

Next, to derive the relation between the physical quantities across the rarefaction wave, we derive the Riemann invariant for the equation of state given by Eq.\,(\ref{EoS}). The Eulerian continuity equation and equation of motion in one dimension are

\begin{align}
& \frac{\partial \rho}{\partial t}+\rho\frac{\partial v}{\partial x}+v\frac{\partial \rho}{\partial x}=0, \label{Euler-EoC} \\ &\frac{\partial v}{\partial t}+v\frac{\partial v}{\partial x}=-\frac{1}{\rho}\frac{\partial P}{\partial x}. \label{Euler-EoM}
\end{align}

\noindent Using Eq.\,(\ref{EoS}), we can transform Eq.\,(\ref{Euler-EoM}) into

\begin{equation}
\frac{\partial v}{\partial t}+v\frac{\partial v}{\partial x}+C_{s}\frac{\partial C_{s} \ln \rho}{\partial x}=0. 
\label{Euler-EoM-2}
\end{equation}

\noindent If we multiply both sides of Eq.\,(\ref{Euler-EoC}) by $C_{s}/\rho$, we obtain

\begin{equation}
\frac{\partial C_{s}\ln \rho}{\partial t}+C_{s}\frac{\partial v}{\partial x}+v\frac{\partial C_{s}\ln \rho}{\partial x}=0.
\label{Euler-EoC-2}
\end{equation}

\noindent Finally, by taking the sum and difference of Eq.\,(\ref{Euler-EoM-2}) and Eq.\,(\ref{Euler-EoC-2}), we obtain

\begin{equation}
\frac{\partial}{\partial t}(v\pm C_{s} \ln \rho )+(v\pm C_{s})\frac{\partial}{\partial x}(v \pm C_{s}\ln \rho )=0.
\label{EoC-EoM}
\end{equation}

Equation (\ref{EoC-EoM}) means the Riemann invariants $J_{\pm}=v\pm C_{s}\ln \rho$ are conserved on a trajectory $dx/dt =v\pm C_{s}$. These Riemann invariants remain constant across the rarefaction wave, thus

\begin{equation}
v\mp C_{s}\ln \rho =v^{\ast} \mp C_{s} \ln \rho^{\ast},
\label{Riemann-invariant-relation}
\end{equation}

\noindent where the sign shows propagation in the opposite direction to the rarefaction wave. If we define $W_{s}$ for the rarefaction wave as in Eq.\,(\ref{Rankine-Hugoniot-momentum}), and use Eq.\,(\ref{Riemann-invariant-relation}) and Eq.\,(\ref{EoS-postshock}), we obtain

\begin{equation}
W_{s} \equiv \left|\frac{P^{\ast}-P}{v^{\ast}-v}\right|=\frac{P-P^{\ast}}{C_{s}}\Bigl[\ln\Bigl( \frac{C_{s}^{2}\rho}{P^{\ast}+C_{s}^{2}\rho_{0,{\rm eos}}} \Bigr)\Bigr]^{-1}.
\label{W-rarefaction}
\end{equation}

We can calculate $P^{\ast}$ and $v^{\ast}$ using Eq.\,(\ref{W-shock}) and Eq.\,(\ref{W-rarefaction}). From the definition of $W_{s}$, we obtain the following relations for the waves that propagate in the right and left directions:

\begin{align}
&-W_{s,{\rm L}}(v^{\ast}-v_{{\rm L}})-(P^{\ast}-P_{{\rm L}})=0, \label{W-P-L} \\ &W_{s,{\rm R}}(v^{\ast}-v_{{\rm R}})-(P^{\ast}-P_{{\rm R}})=0, \label{W-P-R}
\end{align}

\noindent where L and R denote the values on the left-hand side and right-hand side of the initial discontinuity, and

\begin{align}
&W_{s,{\rm D}}=C_{s}\sqrt{\rho_{{\rm D}} \Bigl(\rho_{{\rm D}} + \frac{P^{\ast}-P_{{\rm D}}}{C_{s}^{2}}\Bigr)} \ \ {\rm if} \ \ P^{\ast} \geq P_{{\rm D}}, \nonumber \\ & W_{s,{\rm D}} =\frac{P_{{\rm D}}-P^{\ast}}{C_{s}}\Bigl[\ln \Bigl( \frac{C_{s}^{2}\rho_{{\rm D}}}{P^{\ast}+C_{s}^{2}\rho_{0,{\rm eos}}} \Bigr)\Bigr]^{-1} \ \ {\rm if} \ \ P^{\ast} < P_{{\rm D}} \label{define-W},
\end{align}

\noindent where D represents L or R. Eliminating $v^{\ast}$ or $P^{\ast}$ from Eq.\,(\ref{W-P-L}) and Eq.\,(\ref{W-P-R}), we obtain

\begin{align}
&P^{\ast}=\frac{P_{{\rm L}}/W_{s,{\rm L}}+P_{{\rm R}}/W_{s,{\rm R}}+v_{{\rm L}}-v_{{\rm R}}}{1/W_{s,{\rm L}}+1/W_{s,{\rm R}}}, \label{define-Past} \\&v^{\ast}=\frac{v_{{\rm L}}W_{s,{\rm L}}+v_{{\rm R}}W_{s,{\rm R}}+P_{{\rm L}}-P_{{\rm R}}}{W_{s,{\rm L}}+W_{s,{\rm R}}}. \label{define-vast}
\end{align}

We can determine $P^{\ast}$ and $v^{\ast}$ by iterative calculation of Eq.\,(\ref{define-W}) and Eq.\,(\ref{define-Past}). In practical calculations, 5 cycles of iterations with the initial value $P^{\ast}=(P_{L}+P_{R})/2$ are sufficient.

Many general equations of state, such as the Tillotson equation of state (e.g., \cite{Benz-Asphaug1999}), are composed of a combination of an ideal gas equation of state, a polytropic relation $P=K\rho^{\gamma}$, and the equation of state given by Eq.\,(\ref{EoS}). Therefore we expect that a Riemann solver for general equations of state can be written as a combination of these Riemann solvers.

\section{Result of Linear Stability Analysis}
In this section, we conduct a linear stability analysis of the Godunov SPH method and evaluate its stability against the tensile instability for linear interpolation, cubic spline interpolation, and quintic spline interpolation of $V_{ij}^{2}$.

\subsection{One-dimensional case}
\begin{figure}[!htb]
 \begin{center}
 \includegraphics[width=9cm,height=6cm]{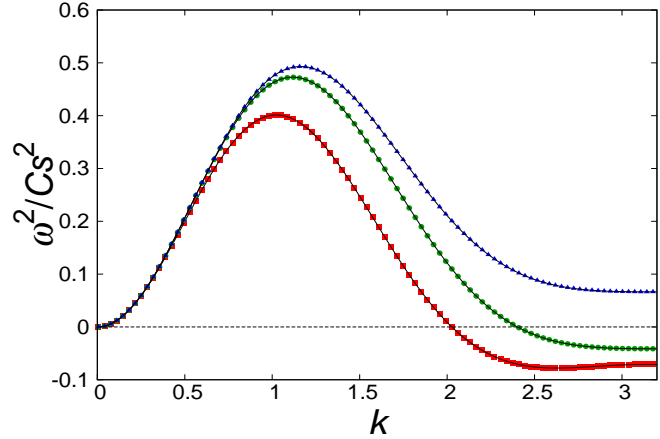}
 \caption{Dispersion relation for a sound wave with negative pressure using linear interpolation (red squares), cubic spline interpolation (green circles) and quintic spline interpolation (blue triangles) in a one-dimensional code. The points show the results obtained using the method shown in Appendix A, and the solid curves show the analytical solution for each interpolation.}
 \label{DR-1D-negative}
 \end{center}
\end{figure}

First, we conduct a linear stability analysis of sound wave propagation using a one-dimensional code. Figure \ref{DR-1D-negative} shows the dispersion relation for linear interpolation, cubic spline interpolation, and quintic spline interpolation in a one-dimensional flow with negative pressure. The parameters are the same as in Fig.\,\ref{DR-numerical-integration} and Fig.\,\ref{DR-numerical-Vij2}: $m=0.0008$, $h=0.001$, $\Delta x=0.001$, $C_{s}=1.0$, $\rho_{0,{\rm eos}}=1.0$, and the average pressure is $P_{0}=-0.2$ (negative).

Each point shows the results that are calculated using the method shown in Appendix A, and the solid curves show the analytical solutions that are obtained from a linear analysis of each interpolation. Here, we only show the formula of the analytical solution, and the details of the linear analysis are shown in Appendix B.

In the case of linear interpolation,

\begin{equation}
\omega^{2}_{{\rm linear}}=-C_{s}^{2}Da+\frac{2DP_{0}}{\rho_{0}}a+\frac{2P_{0}}{\rho_{0}}b, 
\label{DR-linear-1D}
\end{equation}

\noindent where

\begin{align}
&D=\sum_{j}-\sin [k(\overline{x_{i}}-\overline{x_{j}})]\frac{\partial}{\partial \overline{x_{i}}}W(\overline{x_{i}}-\overline{x_{j}},h)\frac{m}{\rho_{0}}, \nonumber \\ &a=\sum_{j\neq i}\sin [k(\overline{x_{i}}-\overline{x_{j}})]\frac{\partial}{\partial \overline{x_{i}}}W(\overline{x_{i}}-\overline{x_{j}},\sqrt{2}h)\frac{m}{\rho_{0}}, \nonumber \\ & b=\sum_{j\neq i}(1-\cos [k(\overline{x_{i}}-\overline{x_{j}})])\frac{\partial^{2}}{\partial \overline{x_{i}}^{2}}W(\overline{x_{i}}-\overline{x_{j}},\sqrt{2}h)\frac{m}{\rho_{0}}. \label{DR-linear-1D-coefficients}
\end{align}

In the case of cubic spline interpolation,

\begin{equation}
\omega_{{\rm cubic}}^{2}=-C_{s}^{2}Da + \frac{2P_{0}}{\rho_{0}}b - \frac{P_{0}}{\rho_{0}}\Bigl[ -2Da + \frac{1}{2}h^{2}C_{\rho}c-\frac{1}{2}C_{\rho}d\Bigr],
\label{DR-cubic-1D}
\end{equation}

\noindent where

\begin{align}
&C_{\rho}=\sum_{j}(1-\cos [k(\overline{x_{i}}-\overline{x_{j}})])\frac{\partial^{2}}{\partial \overline{x_{i}}^{2}}W(\overline{x_{i}}-\overline{x_{j}},h)\frac{m}{\rho_{0}}, \nonumber \\ & c=\sum_{j\neq i}\frac{1}{\overline{x_{i}}-\overline{x_{j}}}(1-\cos [k(\overline{x_{i}}-\overline{x_{j}})])\frac{\partial}{\partial \overline{x_{i}}}W(\overline{x_{i}}-\overline{x_{j}},\sqrt{2}h)\frac{m}{\rho_{0}}, \nonumber \\ & d=\sum_{j\neq i}(\overline{x_{i}}-\overline{x_{j}})(1-\cos [k(\overline{x_{i}}-\overline{x_{j}})])\frac{\partial}{\partial \overline{x_{i}}}W(\overline{x_{i}}-\overline{x_{j}},\sqrt{2}h)\frac{m}{\rho_{0}}. \label{DR-cubic-1D-coefficients}
\end{align}

In the case of quintic spline interpolation,

\begin{align}
\omega_{{\rm quintic}}^{2}=&-C_{s}^{2}DA_{0}+\frac{2P_{0}}{\rho_{0}}C-\frac{P_{0}}{\rho_{0}}\Bigl[ -\frac{3}{8}h^{4}C_{\rho}B_{-3}-\frac{3}{16}h^{4}C_{\rho}DA_{-2}+\frac{3}{4}h^{2}C_{\rho}B_{-1}\nonumber \\ &+\frac{1}{8}h^{2}C_{\rho}DA_{0}-2DA_{0}-\frac{5}{8}C_{\rho}B_{1}-\frac{1}{16}C_{\rho}DA_{2} \Bigr], \label{DR-quintic-1D}
\end{align}

\noindent where

\begin{align}
&A_{n}=\sum_{j\neq i}(\overline{x_{i}}-\overline{x_{j}})^{n}\sin [k(\overline{x_{i}}-\overline{x_{j}})]\frac{\partial}{\partial \overline{x_{i}}}W(\overline{x_{i}}-\overline{x_{j}},\sqrt{2}h)\frac{m}{\rho_{0}}, \nonumber \\ &B_{n}=\sum_{j\neq i}(\overline{x_{i}}-\overline{x_{j}})^{n}(1-\cos [k(\overline{x_{i}}-\overline{x_{j}})])\frac{\partial}{\partial \overline{x_{i}}}W(\overline{x_{i}}-\overline{x_{j}},\sqrt{2}h)\frac{m}{\rho_{0}}, \nonumber \\ &C=\sum_{j\neq i}(1-\cos [k(\overline{x_{i}}-\overline{x_{j}})])\frac{\partial ^{2}}{\partial \overline{x_{i}}^{2}}W(\overline{x_{i}}-\overline{x_{j}},\sqrt{2}h)\frac{m}{\rho_{0}}, \label{DR-quintic-1D-coefficients}
\end{align}

\noindent and $\overline{x_{i}}$ and $\overline{x_{j}}$ show the position of a particle in the unperturbed state.

As we can see in Fig.\,\ref{DR-1D-negative}, the dispersion relations obtained using the method of Appendix A agree with the analytical solutions given by Eq.\,(\ref{DR-linear-1D}), Eq.\,(\ref{DR-cubic-1D}), and Eq.\,(\ref{DR-quintic-1D}). In the cases of linear interpolation and cubic spline interpolation, $\omega^{2}<0$ at short wavelengths (large $k$). Thus these calculations are unstable. However, in the case of quintic spline interpolation, $\omega^{2}>0$ even at short wavelengths, and so this quintic spline interpolation is stable for negative pressure.

These results do not depend on the absolute value of $P_{0}$ or $\rho_{0}$, as long as the pressure is negative, because $a$ in the first term of Eq.\,(\ref{DR-linear-1D}) and Eq.\,(\ref{DR-cubic-1D}), and $A_{0}$ in the first term of Eq.\,(\ref{DR-quintic-1D}), become 0 at the Nyquist frequency ($k=2\pi /2\Delta x$) and all the other terms are proportional to $P_{0}/\rho_{0}$. Therefore, these results might be general as long as the pressure is negative in the one-dimensional case. 

\begin{figure}[!htb]
  \begin{center}
    \includegraphics[width=9cm,height=6cm]{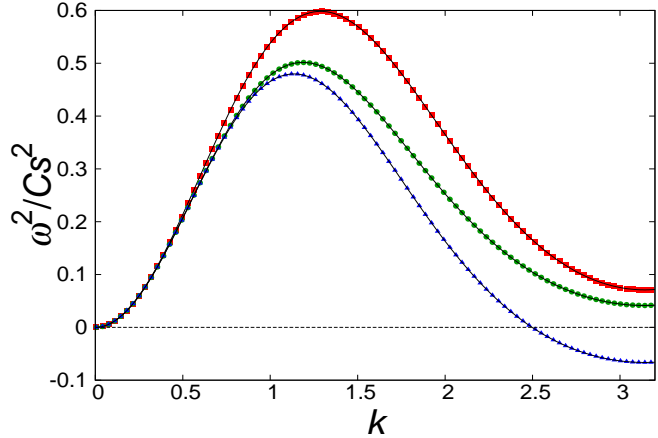}
    \caption{Same as Fig.\,\ref{DR-1D-negative} but for positive pressure.}
    \label{DR-1D-positive}
  \end{center}
\end{figure}

Surprisingly, this fact shows that, at least at the Nyquist frequency, $\omega^{2}>0$ for negative pressure implies $\omega^{2}<0$ for positive pressure. Figure \ref{DR-1D-positive} shows the dispersion relation for the same conditions as in Fig.\,\ref{DR-1D-negative} but with $\rho_{0,{\rm eos}}=0.6$ and hence positive pressure, $P_{0}=0.2$.

As shown in Fig.\,\ref{DR-1D-positive}, linear interpolation and cubic spline interpolation are stable, but quintic spline interpolation is unstable at short wavelengths. Therefore, we conclude that quintic spline interpolation is stable for negative pressure but unstable for positive pressure in the one-dimensional case.

\subsection{Two-dimensional and three-dimensional case}
Next, we conduct a linear stability analysis for the two-dimensional and three-dimensional cases. We put the particles on a Cartesian lattice with a side length of $\Delta x$ in the unperturbed state. For simplicity we assume that the wave number vector is along the $x$-axis. In this case the analytical solutions of the dispersion relation for linear interpolation and cubic spline interpolation are

\begin{align}
& \omega^{2}_{{\rm linear}}=-C_{s}^{2}Da+\frac{2DP_{0}}{\rho_{0}}a+\frac{2P_{0}}{\rho_{0}}b,  \label{DR-linear-2-3D} \\ &\omega_{{\rm cubic}}^{2}=-C_{s}^{2}Da + \frac{2P_{0}}{\rho_{0}}b - \frac{P_{0}}{\rho_{0}}\Bigl[ -2Da + \frac{1}{2}h^{2}C_{\rho}c-\frac{1}{2}C_{\rho}d\Bigr]. \label{DR-cubic-2-3D}
\end{align}

Actually, the forms of Eq.\,(\ref{DR-linear-2-3D}) and Eq.\,(\ref{DR-cubic-2-3D}) appear to be the same as the one-dimensional case, Eq.\,(\ref{DR-linear-1D}) and Eq.\,(\ref{DR-cubic-1D}), but the definitions of the coefficients are different and include summation for the $y$- and $z$-direction:

\begin{align}
& D=\sum_{j}-\sin [k(\overline{x_{i}}-\overline{x_{j}})]\frac{\partial}{\partial \overline{x_{i}}}W(\overline{\bm{r}_{i}}-\overline{\bm{r}_{j}},h)\frac{m}{\rho_{0}}, \nonumber \\ &C_{\rho}=\sum_{j}(1-\cos [k(\overline{x_{i}}-\overline{x_{j}})])\frac{\partial^{2}}{\partial \overline{x_{i}}^{2}}W(\overline{\bm{r}_{i}}-\overline{\bm{r}_{j}},h)\frac{m}{\rho_{0}}, \nonumber \\ &a=\sum_{j\neq i}\sin [k(\overline{x_{i}}-\overline{x_{j}})]\frac{\partial}{\partial \overline{x_{i}}}W(\overline{\bm{r}_{i}}-\overline{\bm{r}_{j}},\sqrt{2}h)\frac{m}{\rho_{0}}, \nonumber \\ &b=\sum_{j\neq i}(1-\cos [k(\overline{x_{i}}-\overline{x_{j}})])\frac{\partial^{2}}{\partial \overline{x_{i}}^{2}}W(\overline{\bm{r}_{i}}-\overline{\bm{r}_{j}},\sqrt{2}h)\frac{m}{\rho_{0}}, \nonumber \\ &c=\sum_{j\neq i}\frac{\overline{x_{i}}-\overline{x_{j}}}{|\overline{\bm{r}_{i}}-\overline{\bm{r}_{j}}|^{2}}(1-\cos [k(\overline{x_{i}}-\overline{x_{j}})])\frac{\partial}{\partial \overline{x_{i}}}W(\overline{\bm{r}_{i}}-\overline{\bm{r}_{j}},\sqrt{2}h)\frac{m}{\rho_{0}}, \nonumber \\ &d=\sum_{j\neq i}(\overline{x_{i}}-\overline{x_{j}})(1-\cos [k(\overline{x_{i}}-\overline{x_{j}})])\frac{\partial}{\partial \overline{x_{i}}}W(\overline{\bm{r}_{i}}-\overline{\bm{r}_{j}},\sqrt{2}h)\frac{m}{\rho_{0}}. \label{DR-linear-cubic-2-3D-coefficients}
\end{align}

Note that the values of the coefficients, apart from $c$, are almost independent of the number of the spatial dimension. The dispersion relation for linear interpolation does not have coefficient $c$, and the results of the two-dimensional and three-dimensional cases are the same as those of the one-dimensional case; linear interpolation is stable for positive pressure and unstable for negative pressure.

The analytical dispersion relation for quintic spline interpolation in the two-dimensional and three-dimensional cases are

\begin{align}
\omega_{{\rm quintic}}^{2}=&-C_{s}^{2}DA_{0,0}+\frac{2P_{0}}{\rho_{0}}C-\frac{P_{0}}{\rho_{0}}\Bigl[ -\frac{3}{8}h^{4}C_{\rho}B_{1,4}-\frac{3}{16}h^{4}C_{\rho}DA_{2,4}+\frac{3}{4}h^{2}C_{\rho}B_{1,2} \nonumber \\ &+\frac{1}{8}h^{2}C_{\rho}DA_{2,2}-2DA_{0,0}-\frac{5}{8}C_{\rho}B_{1,0}-\frac{1}{16}C_{\rho}DA_{2,0} \Bigr], \label{DR-quintic-2-3D}
\end{align}

\noindent where

\begin{align}
& A_{n,m}=\sum_{j}\frac{(\overline{x_{i}}-\overline{x_{j}})^{n}}{|\overline{\bm{r}_{i}}-\overline{\bm{r}_{j}}|^{m}}\sin [k(\overline{x_{i}}-\overline{x_{j}})]\frac{\partial}{\partial \overline{x_{i}}}W(\overline{\bm{r}_{i}}-\overline{\bm{r}_{j}},\sqrt{2}h)\frac{m}{\rho_{0}}, \nonumber \\ & B_{n,m}=\sum_{j}\frac{(\overline{x_{i}}-\overline{x_{j}})^{n}}{|\overline{\bm{r}_{i}}-\overline{\bm{r}_{j}}|^{m}}(1-\cos [k(\overline{x_{i}}-\overline{x_{j}})])\frac{\partial}{\partial \overline{x_{i}}}W(\overline{\bm{r}_{i}}-\overline{\bm{r}_{j}},\sqrt{2}h)\frac{m}{\rho_{0}}, \nonumber \\ & C=\sum_{j}(1-\cos [k(\overline{x_{i}}-\overline{x_{j}})])\frac{\partial^{2}}{\partial \overline{x_{i}}^{2}}W(\overline{\bm{r}_{i}}-\overline{\bm{r}_{j}},\sqrt{2}h)\frac{m}{\rho_{0}}. \label{DR-quintic-2-3D-coefficients}
\end{align}

\begin{figure}[!htb]
  \begin{center}
    \includegraphics[width=9cm,height=6cm]{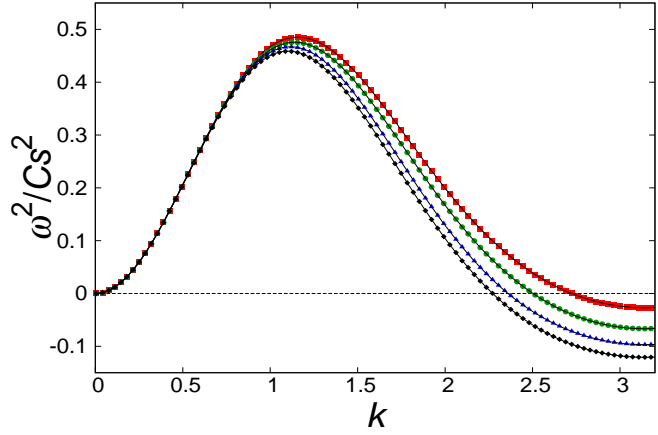}
    \caption{Dispersion relation for cubic spline interpolation in the two-dimensional case (red squares), cubic spline interpolation in the three-dimensional case (green circles), quintic spline interpolation in the two-dimensional case (blue triangles), and quintic spline interpolation in the three-dimensional case (black diamonds) for positive pressure.}
    \label{DR-2-3D-positive}
  \end{center}
\end{figure}

\begin{figure}[!htb]
  \begin{center}
    \includegraphics[width=9cm,height=6cm]{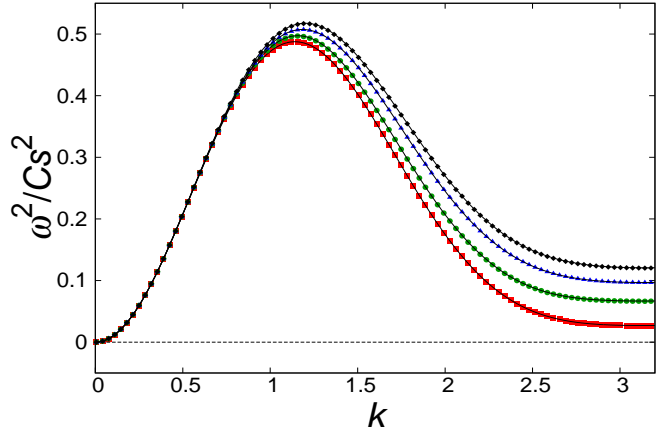}
    \caption{Same as Fig.\,\ref{DR-2-3D-positive} but for negative pressure.}
    \label{DR-2-3D-negative}
  \end{center}
\end{figure}

Figure \ref{DR-2-3D-positive} shows the dispersion relations for cubic spline interpolation and quintic spline interpolation in the two-dimensional and three-dimensional cases for positive pressure, and Fig.\,\ref{DR-2-3D-negative} shows the same dispersion relations but for negative pressure. Here, the parameters are the same as the one-dimensional case (Fig.\,\ref{DR-1D-negative} and Fig.\,\ref{DR-1D-positive}), but to set the average density to $\rho_{0}=0.8$ we set the mass of each particle to $m=0.8\times 10^{-6}$ in the two-dimensional case and $m=0.8\times 10^{-9}$ in the three-dimensional case. 

As we can see in Fig.\,\ref{DR-2-3D-positive} and Fig.\,\ref{DR-2-3D-negative}, quintic spline interpolation has the same result as in the one-dimensional case; it is stable for negative pressure and unstable for positive pressure. However, the correspondence is very different for cubic spline interpolation. Cubic spline interpolation has the opposite result to the one-dimensional case; it is unstable for positive pressure and stable for negative pressure.

This difference between the one-dimensional case and the two- and three-dimensional cases for cubic spline interpolation comes from the coefficient $c$, because all the coefficients, except for $c$, are almost the same for all spatial dimensions. 

\subsection{Variable smoothing length case}
The extension of the method to the variable smoothing length case is not straightforward because Eq.\,(\ref{EoM-variableh}) is not exact in this case. However, we can satisfy almost the same condition as for constant smoothing length in the neighborhood of the $i$-th and $j$-th particles if $C_{{\rm smooth}}$ becomes sufficiently large, because it produces very smooth spatial variation of the smoothing length. Thus we expect that if $C_{{\rm smooth}}$ becomes large we can obtain the same result as for the constant smoothing length case.

Figure \ref{DR-variable-h} shows the dispersion relation calculated using the method of Appendix A in the one-dimensional case, where we use the equation of motion for variable smoothing length, Eq.\,(\ref{EoM-variableh}), quintic spline interpolation, $\rho_{0}=0.8$, $\rho_{0,{\rm eos}}=10.0$, and $P_{0}=-9.2$. The other parameters are the same as in Fig.\,\ref{DR-1D-negative}. The results shown are for $C_{{\rm smooth}}=1.0$, $2.5$, and $5.0$.

\begin{figure}[!htb]
  \begin{center}
    \includegraphics[width=9cm,height=6cm]{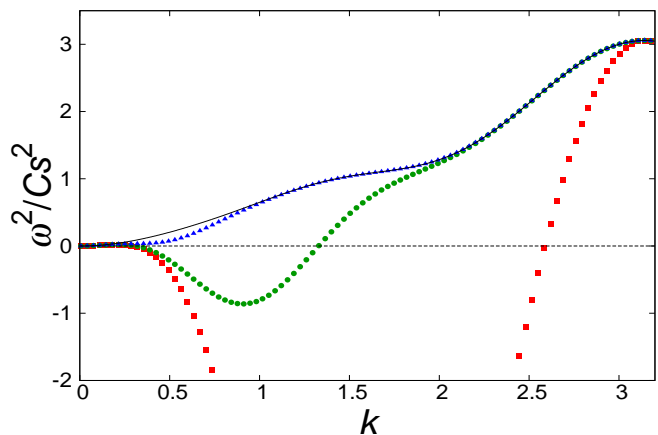}
    \caption{Dispersion relation for variable smoothing length and negative pressure with $\rho_{0,{\rm eos}}=10.0$. Red squares show $C_{{\rm smooth}}=1.0$, green circles show $C_{{\rm smooth}}=2.5$, and blue triangles show $C_{{\rm smooth}}=5.0$. The solid curve shows the analytical dispersion relation for constant smoothing length.}
    \label{DR-variable-h}
  \end{center}
\end{figure}

As shown in Fig.\,\ref{DR-variable-h}, for $C_{{\rm smooth}}=1.0$ the calculation becomes unstable at almost all wavelengths, but for $C_{{\rm smooth}}=5.0$ the calculation is stable at all wavelengths. Moreover, the solid line shows the analytical dispersion relation for constant smoothing length with the same parameters, and we confirm that this solid line almost coincides with the dispersion relation of variable smoothing length with $C_{{\rm smooth}}=5.0$. Therefore, we can obtain the same result as for constant smoothing length if we increase $C_{{\rm smooth}}$. Appropriate values of $C_{{\rm smooth}}$ may depend on the equation of state, the other parameters, the spatial dimension, and so on. We should use sufficiently large $C_{{\rm smooth}}$ in practical calculations.

We should note that even with the variable smoothing length, the tensile instability should not appear if we calculate the exact integration of convolution of EoM (e.g. Eq.\,(77) of \cite{Inutsuka2002}). Therefore, unwanted results for small $C_{{\rm smooth}}$ is due to the poor approximation of the convolutions with the variable smoothing length. We can qualitatively understand why the extension to the variable smoothing length itself aggravates the tensile instability in the negative pressure. The reason is as follows: if two particles approach each other, the density becomes large and the smoothing length becomes short. This makes the shape of the kernel function sharp, and then the force between these two particles becomes strong. Therefore, if the pressure is negative, the attractive force becomes strong, and then this enhances the tensile instability. Large $C_{{\rm smooth}}$ makes this nature of the variable smoothing length weak for the perturbations of short wavelength.

If $C_{{\rm smooth}}$ is larger, the number of neighbor particles required to obtain the correct smoothing length is larger. According to \cite{Inutsuka2002}, the number of neighbor particles for each dimension is $6\eta C_{{\rm smooth}}$ for one dimension, $28\eta^{2}C_{{\rm smooth}}^{2}$ for two dimensions, and $113\eta^{3}C_{{\rm smooth}}^{3}$ for three dimensions. Especially for the two- and three-dimensional cases, the increase of computational cost with larger $C_{{\rm smooth}}$ is significant. However, from Eq.\,(\ref{variable-h}) for the  multidimensional case, the spatial variation of the smoothing length becomes smoother than the one-dimensional case, and then we can consider that the $C_{{\rm smooth}}$ required to obtain the same result as for constant smoothing length becomes smaller.

This approach to variable smoothing length is not ideal. For example, if we improve the derivation of the equation of motion for variable smoothing length, we may formulate a method that works for $C_{{\rm smooth}}=1$. This may emphasize the importance of rigorous formulation of Godunov SPH for variable smoothing length.

\subsection{Summary of results}
Table \ref{table1} shows a summary of stability for the number of dimensions ($d$), various interpolations, and positive and negative pressure in the constant smoothing length case.

\begin{table}[!htb]
  \begin{tabular}{|c||c|c|c|c|} \hline
     & $d=1$, $P>0$ & $d=1$, $P<0$ & $d=2$ or $3$, $P>0$ & $d=2$ or $3$, $P<0$ \\ \hline \hline
    Linear & $\bigcirc$ & $\times$ & $\bigcirc$ & $\times$ \\ \hline
    Cubic & $\bigcirc$ & $\times$ & $\times$ & $\bigcirc$ \\ \hline
    Quintic & $\times$ & $\bigcirc$ & $\times$ & $\bigcirc$ \\ \hline
\end{tabular}
\caption{Summary of stability for the number of dimensions, various interpolations, and positive and negative pressure. A circle ($\bigcirc$) shows stable interpolation and a cross ($\times$) shows unstable interpolation. $d$ represents the spatial dimension and $P$ represents the pressure.}
\label{table1}
\end{table}

According to the test calculation of the shock tube problem in \cite{Inutsuka2002}, the method with linear interpolation is not so accurate at the contact discontinuity. Therefore, it is good to use cubic spline interpolation for positive pressure and quintic spline interpolation for negative pressure in the one-dimensional case. However, for the two- and three-dimensional cases we must use linear interpolation for positive pressure. For negative pressure, both cubic spline interpolation and quintic spline interpolation are stable, but we recommend avoiding quintic spline interpolation because of its heavy computational cost. Thus, in the two- and three-dimensional cases it is good to use linear interpolation for positive pressure and cubic spline interpolation for negative pressure.

To achieve conservation of all momentum and energy, we can use $P_{i}+P_{j}$ as the criterion for the pressure sign. Therefore, the equation of motion in one dimension becomes

\begin{align}
&\dot{\bm{v}}_{i}=-2\sum_{j}m_{j}P_{ij}^{\ast}V_{ij}^{2}\frac{\partial}{\partial \bm{r}_{i}}W(\bm{r}_{i}-\bm{r}_{j},\sqrt{2}h), \nonumber \\ &V_{ij}^{2}=\left\{ \begin{array}{ll} V_{ij,{\rm cubic}}^{2} & {\rm if} \ (P_{i}+P_{j})>0, \\ V_{ij,{\rm quintic}}^{2} & {\rm if} \ (P_{i}+P_{j})<0, \\ \end{array} \right. \label{EoM-for-1D}
\end{align}

\noindent and for two and three dimensions

\begin{align}
 &\dot{\bm{v}}_{i}=-2\sum_{j}m_{j}P_{ij}^{\ast}V_{ij}^{2}\frac{\partial}{\partial \bm{r}_{i}}W(\bm{r}_{i}-\bm{r}_{j},\sqrt{2}h), \nonumber \\ &V_{ij}^{2}=\left\{ \begin{array}{ll} V_{ij,{\rm linear}}^{2} & {\rm if} \ (P_{i}+P_{j})>0, \\ V_{ij,{\rm cubic}}^{2} & {\rm if} \ (P_{i}+P_{j})<0. \\ \end{array} \right. \label{EoM-for-2-3D}
\end{align}

To achieve conservation of energy, we must use the same interpolation of $V_{ij}^{2}$ for the equation of energy. In the case of variable smoothing length, we use Eq.\,(\ref{EoM-variableh}) in the same way as Eq.\,(\ref{EoM-for-1D}) and Eq.\,(\ref{EoM-for-2-3D}), but we should use larger $C_{{\rm smooth}}$ to achieve the same results as in Table \ref{table1}.

\section{Test Calculation}
In this section, we perform test calculations to confirm that the results of the analyses in Section 4 are also valid in actual calculations.

The time integration method is a simple predictor-corrector method as follows: first we calculate the acceleration at the $n$th time step, $\dot{\bm{v}}_{i,n}$, using the physical quantities at the $n$th time step, and then we calculate the time-centered position and velocity as

\begin{align}
&\bm{v}_{i,n+1/2}=\bm{v}_{i,n}+\dot{\bm{v}}_{i,n}\frac{\Delta t}{2}, \nonumber \\ &\bm{r}_{i,n+1/2}=\bm{r}_{i,n}+\bm{v}_{i,n}\frac{\Delta t}{2}+\frac{1}{2}\dot{\bm{v}}_{i,n}\Bigl(\frac{\Delta t}{2}\Bigr)^{2}. \label{time-evolution-1}
\end{align}

\noindent Then we calculate the time-centered acceleration, $\dot{\bm{v}}_{i,n+1/2}$, using the time-centered physical quantities. Finally, we calculate the position and velocity at the next time step as

\begin{align}
&\bm{v}_{i,n+1}=\bm{v}_{i,n}+\dot{\bm{v}}_{i,n+1/2}\Delta t, \nonumber \\ &\bm{r}_{i,n+1}=\bm{r}_{i,n}+\bm{v}_{i,n}\Delta t +\frac{1}{2}\dot{\bm{v}}_{i,n+1/2}\Delta t^{2}. \label{time-evolution-2}
\end{align}

The time step $\Delta t$ is determined by the Courant condition at each time step,

\begin{equation}
\Delta t=\min_{i} C_{{\rm CFL}}\Bigl( \frac{[m_{i}/\rho_{i}]^{1/d}}{C_{s,i}} \Bigr).
\label{Courant-condition}
\end{equation}

\noindent In all the test calculations of this paper we use $C_{{\rm CFL}}=0.5$.

\subsection{Convergence test}
In this subsection, we perform a convergence test to check the accuracy of the Godunov SPH method. We calculate the sound wave in two dimensions as a problem for the convergence test. The density in the unperturbed state is set to 1.0. We use $\rho_{0,{\rm eos}}=0.1$ and $C_{s}=1.0$ for the equation of state. Thus the pressure in the unperturbed state is 0.9. Simulations are performed in the square domain, $x,\ y \in [0.0,1.0]$, and we assume a periodic boundary condition. In the unperturbed state, positions of particles $(\overline{x_{i}},\overline{y_{i}})$ are given as a square lattice. The initial positions and velocities of particles are,

\begin{align}
&x_{i}=\overline{x_{i}}+(0.001/k)\sin (k\overline{x_{i}}), \nonumber \\ & v_{i,x}=-\omega (0.001/k) \cos(k\overline{x_{i}}), \nonumber \\ &y_{i}=\overline{y_{i}}, \nonumber \\ &v_{i,y}=0, \label{initial-for-convergence-test}
\end{align}

\noindent where  $k=\omega =2\pi$. In this case, the amplitude of the density perturbation is 0.001. We consider the case of variable smoothing length with $C_{{\rm smooth}}=1.0$, and use linear interpolation for $V_{ij}^{2}$ because only the linear interpolation is stable for case with the positive pressure in two dimensions.

\begin{figure}[!htb]
  \begin{center}
    \includegraphics[width=9cm,height=6cm]{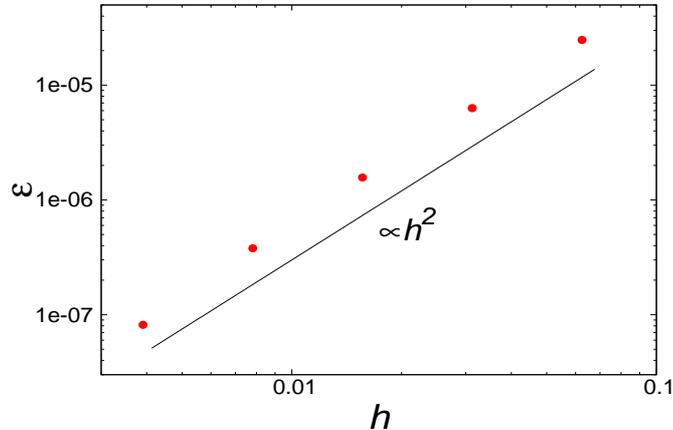}
    \caption{Result of convergence test of the Godunov SPH method for a sound wave. Horizontal axis shows the average smoothing length, and vertical axis shows the relative error $\epsilon$. Solid line represents the line $\propto h^{2}$. Note that the smoothing length is proportional to the average particle spacing.}
    \label{2D-convergence-test}
  \end{center}
\end{figure}

To measure the error, we calculate the difference of the density as,

\begin{equation}
\epsilon = \frac{1}{N_{{\rm tot}}} \sum_{i=1}^{N_{{\rm tot}}}|\rho_{{\rm ref}}(\bm{r}_{i})-\rho_{i}(\bm{r}_{i})|,
\label{error-definition}
\end{equation}

\noindent where $N_{{\rm tot}}$ is the total number of particles, $\rho_{{\rm ref}}(\bm{r}_{i})$ is a reference density at position $\bm{r}_{i}$. In this convergence test, we adopt the result with $N_{{\rm tot}}=512\times 512$ as the reference density. We conduct the calculations with $N_{{\rm tot}}=16\times 16,\ 32\times 32,\ 64\times 64,\ 128\times 128,\ 256\times 256$, and compare with the reference density. To reduce the error due to discrete time integration, time stepping $\Delta t$ is set to very small value $5\times 10^{-4}$ for all resolution. We evaluate the error of density after 100 time-steps. In Fig.\,\ref{2D-convergence-test}, $\epsilon$ is plotted as a function of the average smoothing length, which represents the resolution. 

As we can notice from Fig.\,\ref{2D-convergence-test}, the error is proportional to $h^{2}$. Therefore, it is confirmed that the Godunov SPH method has spatially second-order accuracy.

\subsection{Sound wave in one dimension}
First we calculate the propagation of a sound wave in the one-dimensional case as the simplest test problem.

The particles in the unperturbed state are given equal spacing $\Delta x=0.01$, and $\overline{x_{i}}$ represents the unperturbed position. The initial positions and velocities of particles are given by

\begin{align}
&x_{i}=\overline{x_{i}}+0.01\Delta x\sin(k\overline{x_{i}}), \nonumber \\&v_{i}=-0.01\Delta x\omega \cos(k\overline{x_{i}}), \label{initial-1D-sound-wave}
\end{align}

\noindent where $k$ is the wave number and $\omega$ is the angular frequency of the sound wave. In this calculation we use $k=2\pi$, and to set $C_{s}=\omega /k=1$ we use $\omega =2\pi$. In other words, we resolve one wavelength of the sound wave by 100 particles. We use Eq.\,(\ref{P-inviscid}) for $P_{ij}^{\ast}$ and turn off the dissipation.

All particles have the same mass $m=0.008$, and then the average density is $\rho_{0}=0.8$. We use the equation of state of Eq.\,(\ref{EoS}). In the positive pressure case we use $\rho_{0,{\rm eos}}=0.6$ and the average pressure is $P_{0}=0.2$, and in the negative pressure case we use $\rho_{0,{\rm eos}}=1.0$ and $P_{0}=-0.2$. The boundary condition is periodic.

\begin{figure}[!htb]
  \begin{center}
    \includegraphics[width=9cm,height=6cm]{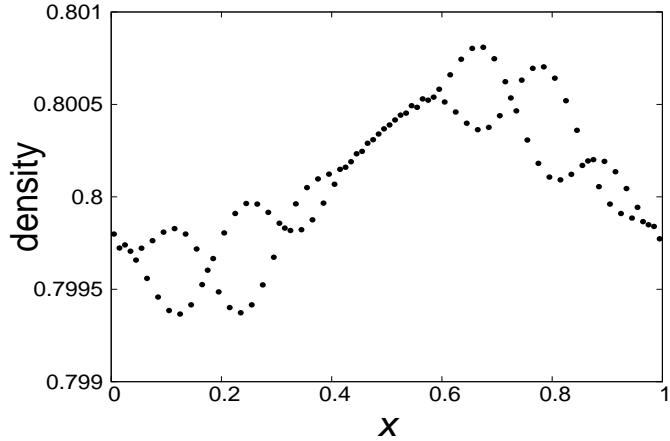}
    \caption{Density distribution of the sound wave using cubic spline interpolation and a constant smoothing length for negative pressure at $t=1.154972$.}
    \label{1D-sound-wave-cubic}
  \end{center}
\end{figure}

\begin{figure}[!htb]
  \begin{center}
    \includegraphics[width=9cm,height=6cm]{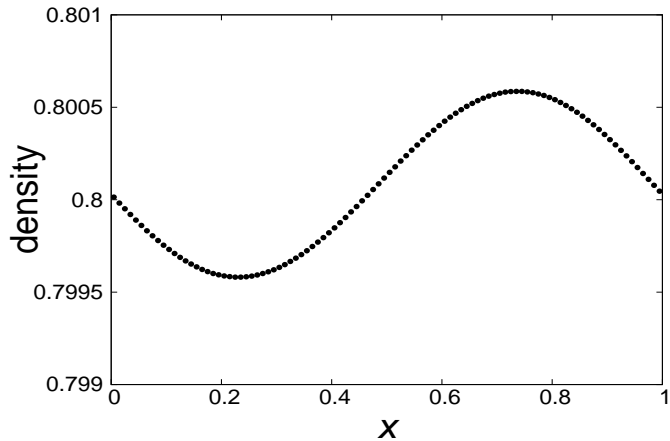}
    \caption{Density distribution of the sound wave using quintic spline interpolation and a constant smoothing length for negative pressure at $t=5.249869$.}
    \label{1D-sound-wave-quintic}
  \end{center}
\end{figure}

Figure \ref{1D-sound-wave-cubic} shows the density distribution of the sound wave with negative pressure using cubic spline interpolation and a constant smoothing length $h=0.01$ immediately after the instability becomes noticeable ($t=1.154972$). Figure \ref{1D-sound-wave-quintic} shows the density distribution for the same conditions but using quintic spline interpolation at $t=5.249869$.

As we can see in Fig.\,\ref{1D-sound-wave-cubic} and Fig.\,\ref{1D-sound-wave-quintic}, our calculations remains stable in the case of quintic spline interpolation even at five times longer than the time at which the calculation with cubic spline interpolation becomes unstable. The calculation with quintic spline interpolation remains stable even at $t=50.0$.

In order to quantify the stability of these calculations, we calculate the time evolution of the minimum distance between all pairs of particles. We define this quantity as

\begin{equation}
r_{{\rm min}}\equiv \min_{{\rm all \ pairs \ of \ }ij}|\bm{r}_{i}-\bm{r}_{j}|/\Delta x. 
\label{define-rmin}
\end{equation}

\begin{figure}[!htb]
  \begin{center}
    \includegraphics[width=9cm,height=6cm]{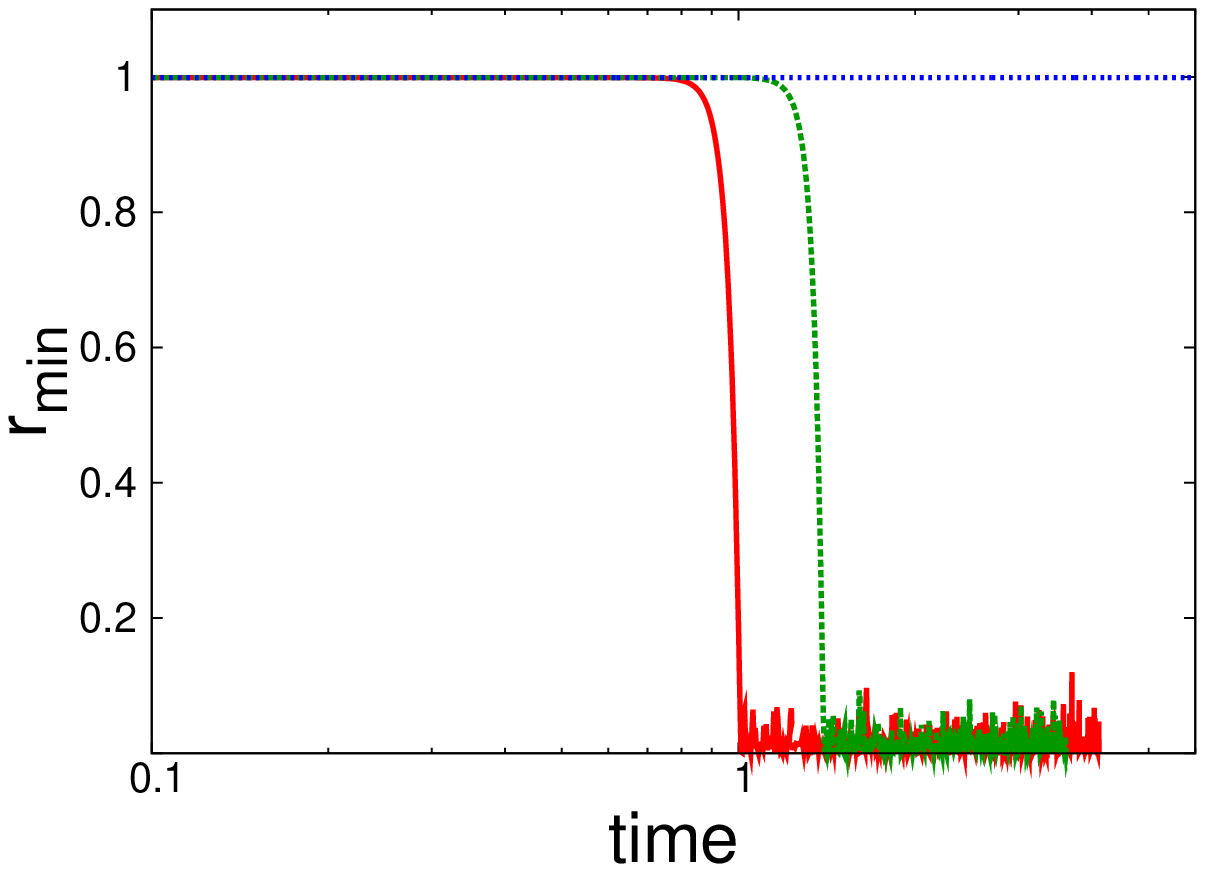}
    \caption{Time evolution of $r_{{\rm min}}$, the minimum distance between adjacent particles, in the case of linear interpolation (red solid line), cubic spline interpolation (green dashed line), and quintic spline interpolation (blue dotted line) during the calculation of the sound wave for negative pressure.}
    \label{rmin-1D-sound-wave-negative}
  \end{center}
\end{figure}

\begin{figure}[!htb]
  \begin{center}
    \includegraphics[width=9cm,height=6cm]{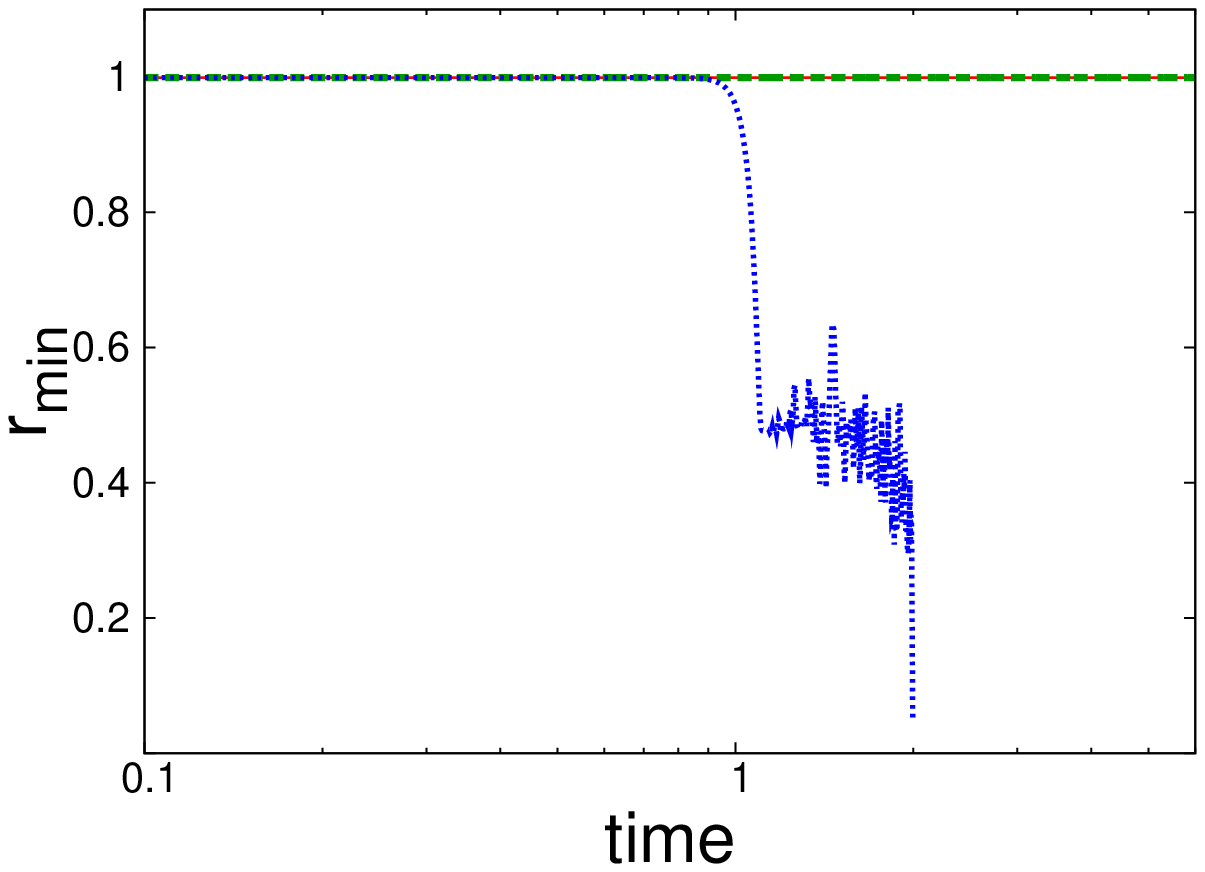}
    \caption{Same as Fig.\,\ref{rmin-1D-sound-wave-negative} but for positive pressure.}
    \label{rmin-1D-sound-wave-positive}
  \end{center}
\end{figure}

\noindent If the calculation is stable, $r_{{\rm min}}$ remains about unity because the distance to the nearest neighbor particle is about $\Delta x$. In contrast, if the particles begin to clump, $r_{{\rm min}}$ becomes small. Figure \ref{rmin-1D-sound-wave-negative} shows the time evolution of $r_{{\rm min}}$ in the case of linear interpolation, cubic spline interpolation, and quintic spline interpolation during the calculation of the negative pressure sound wave, and Fig.\,\ref{rmin-1D-sound-wave-positive} shows the same result but for positive pressure.

As shown in Fig.\,\ref{rmin-1D-sound-wave-negative} and Fig.\,\ref{rmin-1D-sound-wave-positive}, the calculation for negative pressure with quintic spline interpolation remains stable, but for positive pressure the calculations with linear interpolation and cubic spline interpolation remain stable. These results are the same as those of Section 4. Here the lines of unstable calculations are broken because these calculations are terminated.

\begin{figure}[!htb]
  \begin{center}
    \includegraphics[width=9cm,height=6cm]{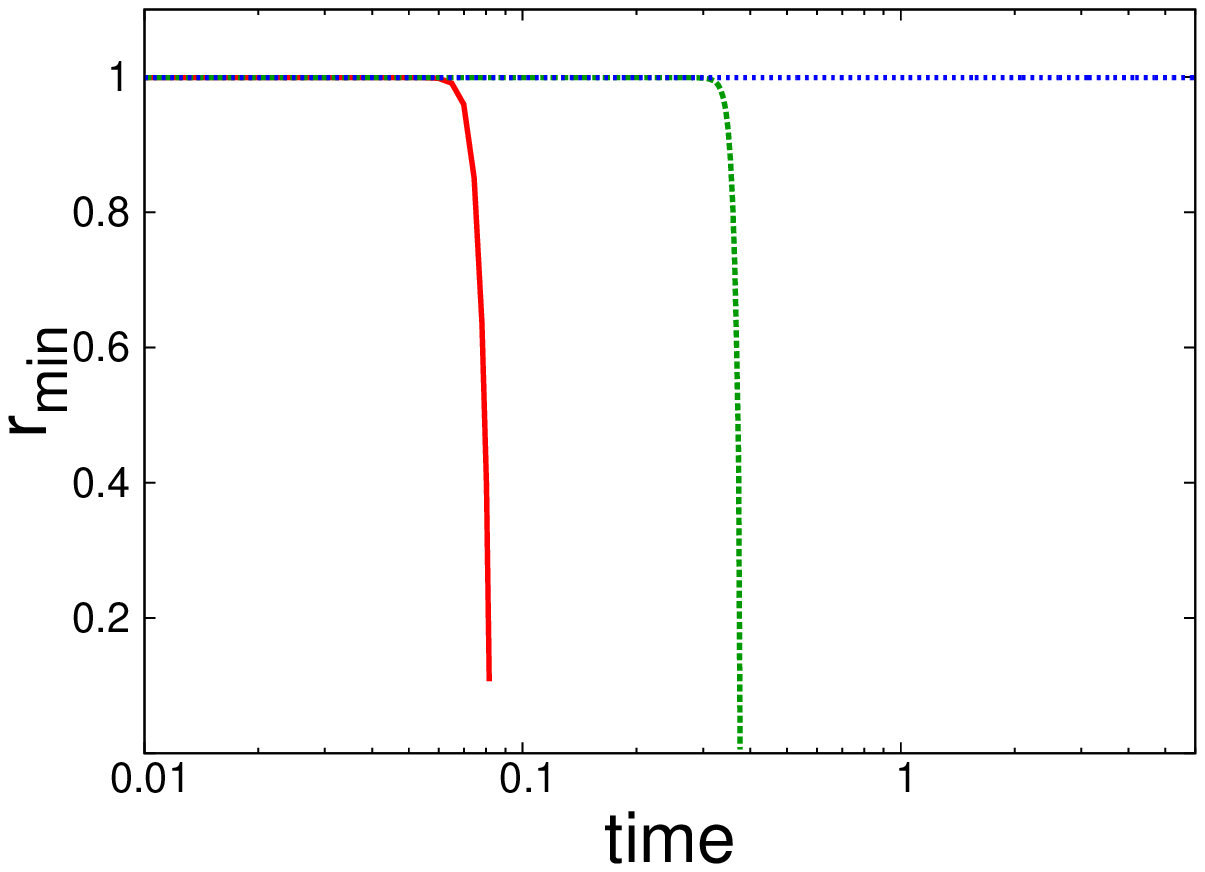}
    \caption{Time evolution of $r_{{\rm min}}$ for the case of variable smoothing length with $C_{{\rm smooth}}=1.0$ (red solid line), $C_{{\rm smooth}}=2.5$ (green dashed line), and $C_{{\rm smooth}}=5.0$ (blue dotted line).}
    \label{rmin-variable-h}
  \end{center}
\end{figure}

Next, we consider the case of variable smoothing length. As in Section 4, we use $\rho_{0,{\rm eos}}=10.0$, $P_{0}=-9.2$, and $\eta=1.0$. The other parameters and initial conditions are the same as in the case of constant smoothing length. Figure \ref{rmin-variable-h} shows the time evolution of $r_{{\rm min}}$ when $C_{{\rm smooth}}=1.0,2.5$, and $5.0$.

As shown in Fig.\,\ref{rmin-variable-h}, in the calculations with $C_{{\rm smooth}}=1.0$ and $C_{{\rm smooth}}=2.5$ the particles stick together soon after the beginning of the calculation. However, with $C_{{\rm smooth}}=5.0$ our calculation is stable. These results also agree with the linear stability analyses. Therefore, stable calculation is possible if we make $C_{{\rm smooth}}$ sufficiently large in the actual calculation with variable smoothing lengths.

\begin{figure}[!htb]
  \begin{center}
    \includegraphics[width=9cm,height=6cm]{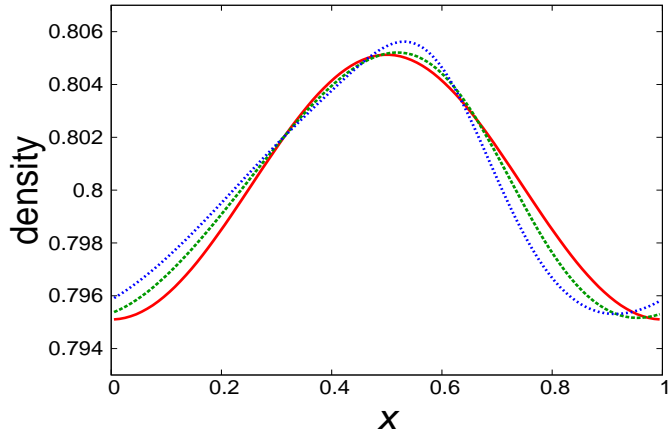}
    \caption{Evolution of the density distribution of the sound wave with negative pressure. The red solid curve shows the density distribution at $t=0.0$, the green dashed curve shows $t=5.0$, and the blue dotted curve shows $t=10.0$.}
    \label{steepening-of-wave}
  \end{center}
\end{figure}

Next, to show the steepening of the wave with high amplitude in the negative pressure case, we start from the initial conditions defined as

\begin{align}
x_{i}=\overline{x_{i}}+0.1\Delta x\sin (k\overline{x_{i}}), \nonumber \\ v_{i}=-0.1 \Delta x\omega \cos (k\overline{x_{i}}), \label{initial-steepening-wave}
\end{align}

\noindent where all parameters are the same as for the case of Fig.\,\ref{1D-sound-wave-quintic}, and we use quintic spline interpolation and a constant smoothing length. Figure \ref{steepening-of-wave} shows the density distribution at $t=0.0$, $t=5.0$, and $t=10.0$.

As we can see from Fig.\,\ref{steepening-of-wave}, the wave profile shows gradual steepening as in the case of positive pressure. Therefore, even in a negative pressure medium, nonlinear steepening of a sound wave is expected to generate shock waves.

\subsection{Shock tube problem}
The calculation of a wave is very simple and physical quantities do not vary largely. Thus, we should test our method for more nonlinear processes, such as those involving a discontinuity. In this subsection, we consider a shock tube problem in a negative pressure medium. To handle the shock wave we need small but finite physical viscosity. In the Godunov method, we use the Riemann solver to introduce a minimum but sufficient viscosity. Therefore, we use the analytical solution of the Riemann solver for Eq.\,(\ref{EoS}) that is derived in Section 3.3 to obtain $P_{ij}^{\ast}$. 

In \cite{Inutsuka2002}, Inutsuka constructed a second-order Riemann solver by considering the spatial gradients of the physical quantities. We also use this second-order Riemann solver. The gradients of the physical quantites are calculated by Eq.\,(\ref{dif-f-of-i}). 

However, in the negative pressure region, the gradient of pressure for perturbation of Nyquist frequency can not be calculated correctly. Using Eq.\,(\ref{dif-f-of-i}), the gradients of density and pressure are calculated as,

\begin{align}
& \nabla \rho_{i}=\sum_{j}m_{j}\frac{\partial}{\partial \bm{r}_{i}}W(\bm{r}_{i}-\bm{r}_{j},h), \label{grad-rho} \\ & \nabla P_{i}=\sum_{j}m_{j}\frac{P_{j}}{\rho_{j}}\frac{\partial}{\partial \bm{r}_{i}}W(\bm{r}_{i}-\bm{r}_{j},h). \label{grad-P}
\end{align}

Here, if we assume Nyquist frequency perturbation and equal mass particles, the density of all particles are the same, and the pressure of all particles are also the same in the case of equation of state Eq.\,(\ref{EoS}). In that case, the gradients of density and pressure become,

\begin{align}
& \nabla \rho_{i}=m\sum_{j}\frac{\partial}{\partial \bm{r}_{i}}W(\bm{r}_{i}-\bm{r}_{j},h), \label{grad-rho-Nyquist} \\ & \nabla P_{i}=m\frac{P}{\rho}\sum_{j}\frac{\partial}{\partial \bm{r}_{i}}W(\bm{r}_{i}-\bm{r}_{j},h), \label{grad-P-Nyquist}
\end{align}

\noindent where, $m$,\ $P$ and $\rho$ are mass, pressure and density respectively. Thus, if $P<0$, the gradients of pressure and density are anti-parallel. This is unphysical, because $\nabla P=C_{s}^{2}\nabla \rho$ with positive $C_{s}^{2}$ means that two gradients are parallel. 

If we use inconsistent gradient of pressure, we tend to estimate the resultant pressure of the Riemann problem for approaching particles mistakenly smaller, and this makes attractive force stronger. Therefore, we should calculate the gradient of pressure by a much better method. One possible way is,

\begin{equation}
\nabla P_{i}=C_{s,i}^{2}\nabla \rho_{i},
\label{modified-grad-P}
\end{equation}

\noindent where, $C_{s,i}$ is local sound speed of $i$-th particle, and $\nabla \rho_{i}$ is calculated by Eq.\,(\ref{grad-rho}). In this paper, we use this modified gradient of pressure. In the case of variable smoothing length, we replace $h$ of Eq.\,(\ref{grad-rho}) with $h_{i}$.

Here, we slightly modify the monotonicity constraint. We use a first-order Riemann solver when there are some particles that have opposite-sign gradients within their neighborhood. This condition is written for pairs of $i$-th and $j$-th particles as

\begin{equation}
\Bigl( \frac{\partial f}{\partial s} \Bigr)_{i} \cdot \Bigl( \frac{\partial f}{\partial s} \Bigr)_{j}<0, 
\label{monotonicity-constraint}
\end{equation}

\noindent where

\begin{align}
&\Bigl( \frac{\partial f}{\partial s} \Bigr)_{i}=\Bigl(\frac{\bm{r}_{i}-\bm{r}_{j}}{|\bm{r}_{i}-\bm{r}_{j}|} \Bigr) \cdot \nabla f_{i}, \nonumber \\ &\Bigl( \frac{\partial f}{\partial s} \Bigr)_{j}=\Bigl(\frac{\bm{r}_{i}-\bm{r}_{j}}{|\bm{r}_{i}-\bm{r}_{j}|} \Bigr) \cdot \nabla f_{j}, \label{define-dif-f-s}
\end{align}

\noindent and $f$ represents $\rho$ or $P$. In this subsection, if there is any one particle $j$ that satisfies the condition Eq.\,(\ref{monotonicity-constraint}) within 3$h_{i}$ from the $i$-th particle, we use the first-order Riemann solver for the $i$-th particle. 

We use the equation of state given by Eq.\,(\ref{EoS}), $C_{s}=1.0$, and $\rho_{0,{\rm eos}}=2.5$. The initial discontinuity of the shock tube problem is at $x=0$, and the initial parameters are

\begin{align}
&\rho_{{\rm L}}=2.0,\rho_{{\rm R}}=1.0, \nonumber \\ &P_{{\rm L}}=-0.5,P_{{\rm R}}=-1.5, \nonumber \\ &v_{x,{\rm L}}=0.0,v_{x,{\rm R}}=0.0, \label{initial-shocktube}
\end{align}

\noindent where L denotes the physical quantities on the left-hand side of the initial discontinuity, and R denotes the right-hand side. The mass of each particle is $m=0.01$, the particle spacing on the left-hand side is $\Delta x_{{\rm L}}=0.005$, and on the right-hand side it is $\Delta x_{{\rm R}}=0.01$. We put 200 particles on the left-hand side and 100 particles on the right-hand side. The boundary condition is a wall boundary ($v_{x}(x=-1)=v_{x}(x=1)=0$).

\begin{figure}[!htb]
  \begin{center}
    \includegraphics[width=9cm,height=6cm]{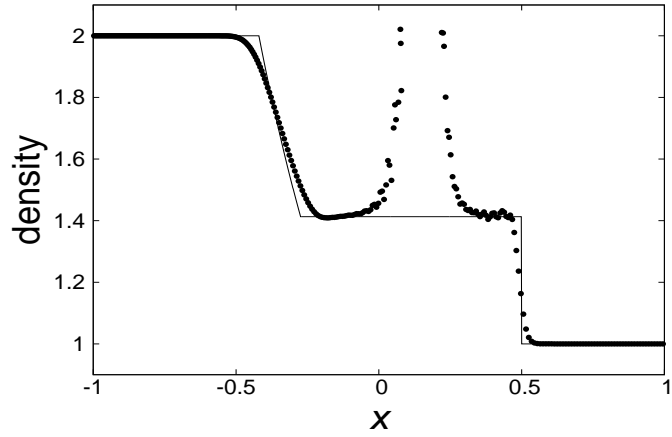}
    \caption{Density distribution of the shock tube problem with negative pressure at $t=0.4200$ obtained by cubic spline interpolation. The solid line corresponds to the analytical solution.}
    \label{shocktube-cubic}
  \end{center}
\end{figure}

\begin{figure}[!htb]
  \begin{center}
    \includegraphics[width=9cm,height=6cm]{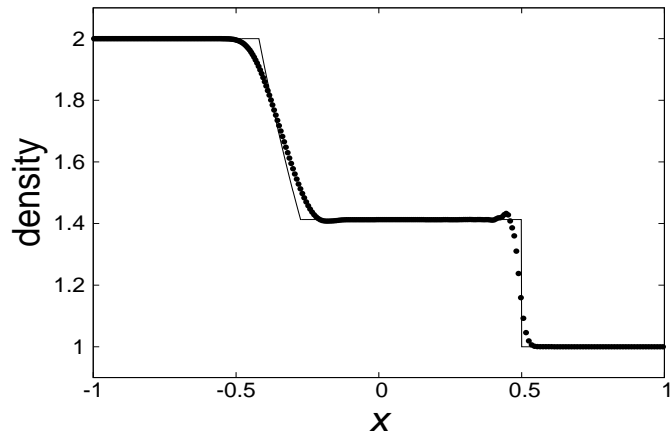}
    \caption{Density distribution of the shock tube problem with negative pressure at $t=0.4200$ obtained by quintic spline interpolation. The solid line corresponds to the analytical solution.}
    \label{shocktube-quintic}
  \end{center}
\end{figure}

First, we compare the results of cubic spline interpolation and quintic spline interpolation. Here we use a constant smoothing length $h=0.01$. Figure \ref{shocktube-cubic} shows the density distribution in the case of cubic spline interpolation immediately after the instability becomes visible ($t=0.4200$), and Fig.\,\ref{shocktube-quintic} shows the density distribution in the case of quintic spline interpolation at $t=0.4200$. Here, the solid curve corresponds to the analytical solution of the shock tube problem that is derived with the analytical solution of the Riemann problem in Section 3.3.

As shown in Fig.\,\ref{shocktube-cubic} and Fig.\,\ref{shocktube-quintic}, in the case of cubic spline interpolation the numerical instability occurs at the initial discontinuity, but in the case of quintic spline interpolation there is no instability. Note that the contact discontinuity does not exist because we use Eq.\,(\ref{EoS}) for the equation of state and the pressure only depends on the density. The result of this simulation matches the analytical solution well. 

\begin{figure}[!htb]
  \begin{center}
    \includegraphics[width=9cm,height=6cm]{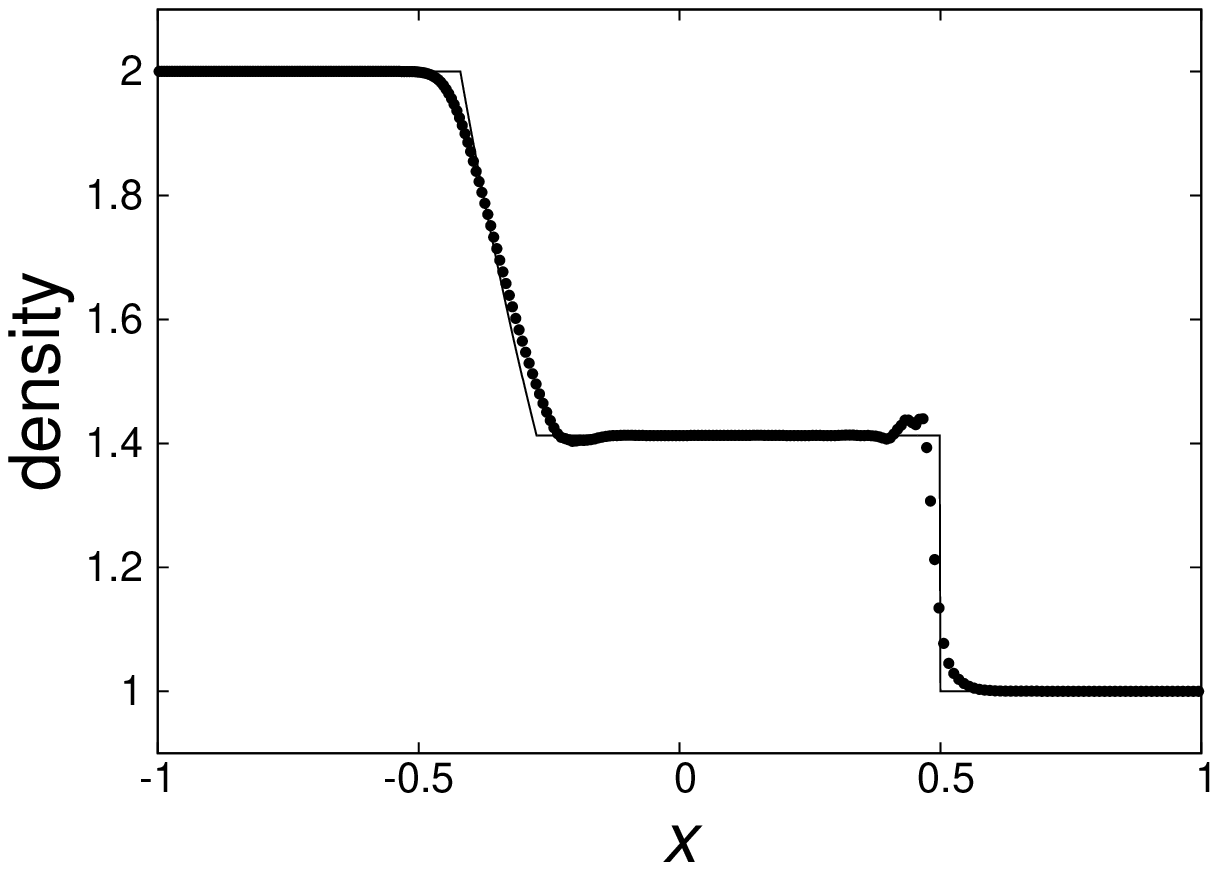}
    \caption{Density distribution at $t=0.4200$ obtained by quintic spline interpolation and variable smoothing length with $C_{{\rm smooth}}=2.0$.}
    \label{shocktube-Csmooth=2.0}
  \end{center}
\end{figure}

Next, we test the calculation with variable smoothing length. Fig.\,\ref{shocktube-Csmooth=2.0} shows the density distribution with $C_{{\rm smooth}}=2.0$ at $t=0.4200$. Here we use $\eta =1.0$.

Our test calculations show that the calculation with $C_{{\rm smooth}}=1.0$ becomes unstable, but the calculation with $C_{{\rm smooth}}=2.0$ is stable even with variable smoothing length. 

Therefore, the stability of the shock tube problem also agrees with the result of the linear stability analysis of a sound wave.

\subsection{Sound wave in two dimensions}
In this subsection, we consider a sound wave in the two-dimensional case. In the unperturbed state, the particles are put on a square lattice with a side length of $\Delta x=0.01$. The initial positions and velocities are

\begin{align}
&x_{i}=\overline{x_{i}}+0.01\Delta x\sin(k\overline{x_{i}}), \nonumber \\ &v_{i,x}=-0.01\Delta x\omega \cos(k\overline{x_{i}}), \nonumber \\ &y_{i}=\overline{y_{i}}, \nonumber \\ &v_{i,y}=0. \label{initial-2D-sound-wave}
\end{align}

In this problem, we consider a sound wave that propagates toward the $x$-direction. We use $k=2\pi /(40\Delta x)=5.0\pi$ and $\omega=5.0\pi$ to set $C_{s}=\omega /k=1.0$. Thus we resolve 1 wavelength with 40 particles. The mass of each particle is $m=0.8\times 10^{-4}$ and the average density is $\rho_{0}=0.8$. The equation of state is again given by Eq.\,(\ref{EoS}), and in the positive pressure case we use $\rho_{0,{\rm eos}}=0.6$ and $P_{0}=0.2$, and in the negative pressure case we use $\rho_{0,{\rm eos}}=1.0$ and $P_{0}=-0.2$. We use Eq.\,(\ref{P-inviscid}) for $P_{ij}^{\ast}$. The boundary condition is periodic for both the $x$- and $y$-directions. In this calculation we use a constant smoothing length $h=0.01$.

\begin{figure}[!htb]
  \begin{center}
    \includegraphics[width=9cm,height=6cm]{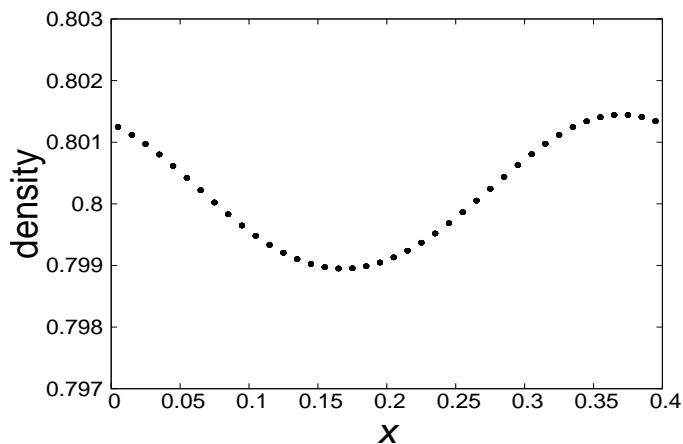}
    \caption{Density distribution in the $x$-direction for two-dimensional sound wave propagation with negative pressure at $t=3.004534$ using cubic spline interpolation.}
    \label{2D-sound-wave-cubic}
  \end{center}
\end{figure}

Figure \ref{2D-sound-wave-cubic} shows the density distribution for the $x$-direction wave with negative pressure using cubic spline interpolation at $t=3.004534$. In the two-dimensional case with negative pressure, the calculation using linear interpolation is unstable. However, as we can see in Fig.\,\ref{2D-sound-wave-cubic}, the calculation with cubic spline interpolation remains stable, and we confirm that this remains stable at least until $t=50.0$.

\begin{figure}[!htb]
  \begin{center}
    \includegraphics[width=9cm,height=6cm]{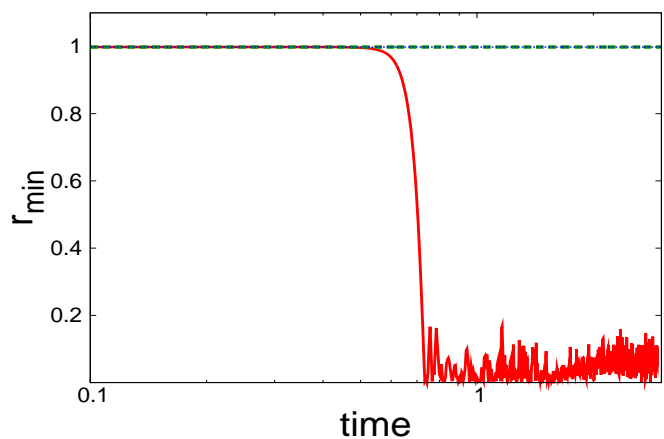}
    \caption{Time evolution of $r_{{\rm min}}$ during the calculation of two-dimensional sound wave propagation with negative pressure.The red solid line shows linear interpolation, the green dashed line shows cubic spline interpolation, and the blue dotted line shows quintic spline interpolation.}
    \label{rmin-2D-sound-wave-negative}
  \end{center}
\end{figure}

\begin{figure}[!htb]
  \begin{center}
    \includegraphics[width=9cm,height=6cm]{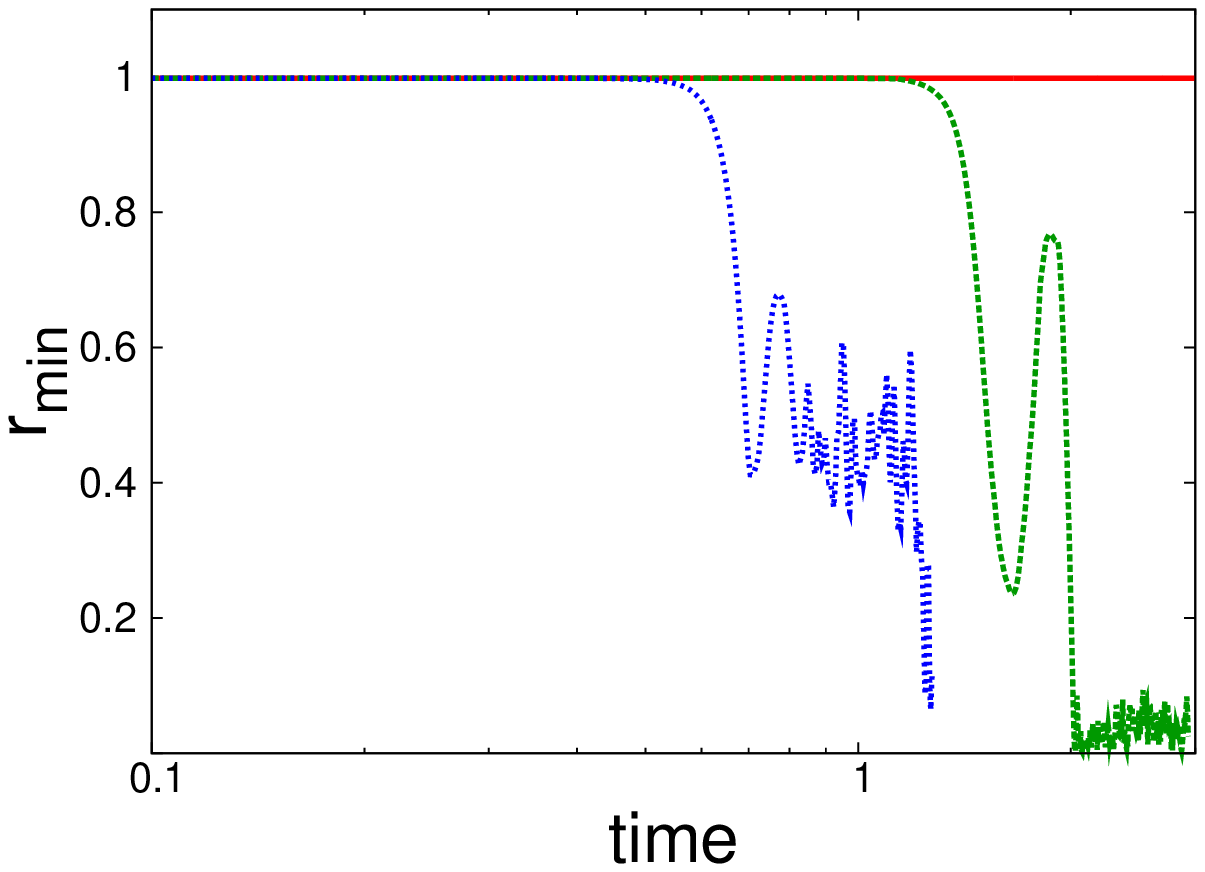}
    \caption{Same as Fig.\,\ref{rmin-2D-sound-wave-negative} but for positive pressure.}
    \label{rmin-2D-sound-wave-positive}
  \end{center}
\end{figure}

Figure \ref{rmin-2D-sound-wave-negative} shows the evolution of $r_{{\rm min}}$ during the calculation with negative pressure using linear interpolation, cubic spline interpolation, and quintic spline interpolation, and Fig.\,\ref{rmin-2D-sound-wave-positive} shows with positive pressure. We can see that $r_{{\rm min}}$ becomes small in the calculation with linear interpolation for negative pressure, and cubic spline interpolation and quintic spline interpolation for positive pressure. The result for cubic spline interpolation is in contrast to the one-dimensional case. These results also agree with the results of the linear stability analyses of a sound wave.

\begin{figure}[!htb]
  \begin{center}
    \includegraphics[width=9cm,height=6cm]{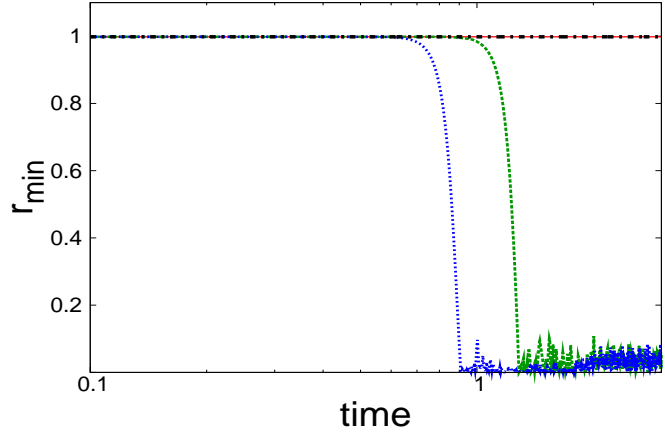}
    \caption{Time evolution of $r_{{\rm min}}$ during the calculation of two-dimensional sound wave propagation. In the unperturbed state, particles are put on a regular triangular lattice. The red solid line shows linear interpolation with positive pressure, the green dashed line shows cubic spline interpolation with positive pressure, the blue dotted line shows linear interpolation with negative pressure, and the black chain line shows cubic spline interpolation with negative pressure.}
    \label{2D-wave-rmin-log-triangle}
  \end{center}
\end{figure}

The Cartesian lattice is not an ideal configuration compared to more stable configuration like the densest-sphere packing. Thus we conduct the same test for the case of a regular triangular lattice. All conditions other than the position in the unperturbed state are the same as the previous test. For simplicity, we test linear interpolation and cubic spline interpolation, and we do not test quintic spline interpolation, because the stability of cubic and quintic spline interpolation is the same in two dimensions. Figure \ref{2D-wave-rmin-log-triangle} shows the evolution of $r_{{\rm min}}$ using linear interpolation and cubic spline interpolation.

As shown in Fig.\,\ref{2D-wave-rmin-log-triangle}, the calculations with linear interpolation for positive pressure and cubic spline interpolation for negative pressure are stable,  but the calculations with linear interpolation for negative pressure and cubic spline interpolation for positive pressure are unstable. These results are consistent with the case of the Cartesian lattice.

Moreover, we conduct the calculation of the sound wave with different direction of the wave number vector. In the unperturbed state, the particles are put on a Cartesian lattice. We use $\bm{k}=(5.0\pi ,5.0\pi)$ and $\omega = |\bm{k}| = 5.0\sqrt{2}\pi$. The initial positions and the velocities are,

\begin{align}
& \bm{r}_{i}=\overline{\bm{r}_{i}}+\delta \bm{r}_{i}, \nonumber \\ & \delta \bm{r}_{i} = (0.01\Delta x,0.01\Delta x)\sin (\bm{k}\cdot \overline{\bm{r}_{i}}), \nonumber \\ & \bm{v}_{i}=-\omega (0.01\Delta x,0.01\Delta x)\cos (\bm{k}\cdot \overline{\bm{r}_{i}}). \label{initial-2D-sound-wave-oblique-direction}
\end{align}

\begin{figure}[!htb]
  \begin{center}
    \includegraphics[width=9cm,height=6cm]{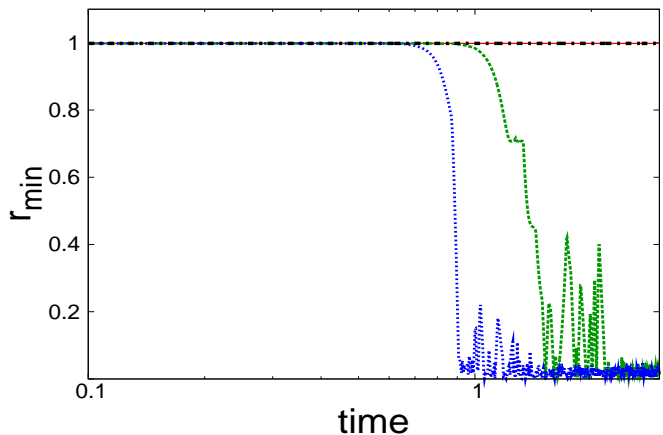}
    \caption{Time evolution of $r_{{\rm min}}$ during the calculation of two-dimensional sound wave propagation with $\bm{k}\propto (1,1)$. The red solid line shows linear interpolation with positive pressure, the green dashed line shows cubic spline interpolation with positive pressure, the blue dotted line shows linear interpolation with negative pressure, and the black chain line shows cubic spline interpolation with negative pressure.}
    \label{2D-wave-rmin-log-oblique-direction}
  \end{center}
\end{figure}

The other parameters and conditions are the same as in the case with the wave number vector is along the $x$-direction. Figure \ref{2D-wave-rmin-log-oblique-direction} shows the evolution of $r_{{\rm min}}$ in this case using linear interpolation and cubic spline interpolation.

As we can notice from Fig.\,\ref{2D-wave-rmin-log-oblique-direction}, the results are the same with those of the lattice-grid aligned cases. 

\subsection{Shock tube problem in two dimensions}
In this subsection, we consider a shock tube problem in two dimensions. To demonstrate the validity of our method in a situation where the sign of pressure changes spatially and temporally, we calculate a shock tube problem with different signs of pressure in left and right side of initial discontinuity. 

Also in this subsection, we use the equation of state given by Eq.\,(\ref{EoS}), $C_{s}=1.0$ and $\rho_{0,{\rm eos}}=2.5$. We set the initial discontinuity at $x=0$, and the region for $x<0$ corresponds to the left side of the initial discontinuity, $x>0$ corresponds to the right-hand side. The initial parameters are,

\begin{align}
&\rho_{{\rm L}}=4.0, \rho_{{\rm R}}=1.0, \nonumber \\ &P_{{\rm L}}=1.5, P_{{\rm R}}=-1.5, \nonumber \\ &v_{x,{\rm L}}=v_{y,{\rm L}}=0.0, v_{x,{\rm R}}=v_{y,{\rm R}}=0.0. \label{initial-2D-shocktube}
\end{align}

We put SPH particles on a regular triangular lattice. The side length of this regular triangle is, 0.01 for the left side and 0.02 for the right side. The mass of each particle is $m=2\sqrt{3} \times 10^{-4}$. In this subsection, we use constant smoothing length with $h=0.02$ for simplicity. The boundary condition for the $x$-direction is a wall boundary ($v_{x}(x=-1)=v_{x}(x=1)=0$), and for the $y$-direction we use a periodic boundary. In the same way as in one dimension, we use the second-order Riemann solver for elastic equation of state.

\begin{figure}[!htb]
  \begin{center}
    \includegraphics[width=9cm,height=6cm]{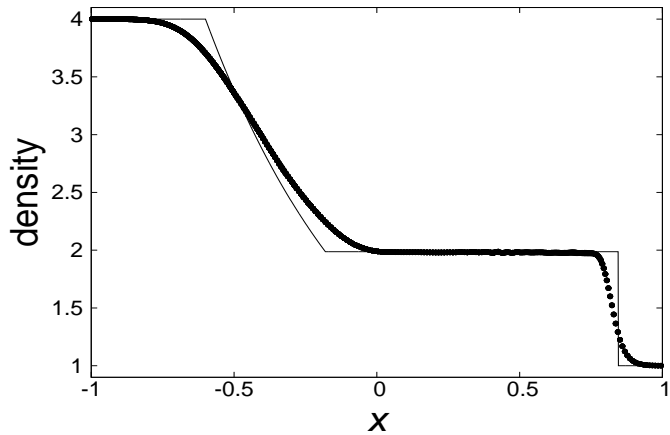}
    \caption{Density distribution of the shock tube problem in two dimensions at $t=0.6000$ obtained by choosing appropriate interpolation. The solid line corresponds to the analytical solution.}
    \label{2D-elastic-shocktube-proper}
  \end{center}
\end{figure}

\begin{figure}[!htb]
  \begin{center}
    \includegraphics[width=9cm,height=6cm]{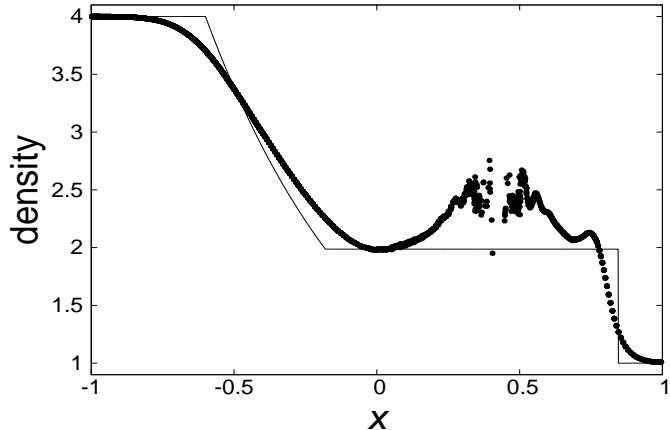}
    \caption{Same as Fig.\,\ref{2D-elastic-shocktube-proper}, but we use only linear interpolation.}
    \label{2D-elastic-shocktube-linear}
  \end{center}
\end{figure}

Figure \ref{2D-elastic-shocktube-proper} shows the density distribution at $t=0.6000$ when we use appropriate interpolation of $V_{ij}^{2}$ as Eq.\,(\ref{EoM-for-2-3D}), and Fig.\,\ref{2D-elastic-shocktube-linear} shows the same density distribution at the same time but we use only linear interpolation independent of the sign of pressure.

As shown in Fig.\,\ref{2D-elastic-shocktube-proper}, if we use appropriate interpolation depending on the sign of pressure, we can calculate without any problem. However, as shown in Fig.\,\ref{2D-elastic-shocktube-linear}, if we ignore the sign of pressure and we use only linear interpolation, the particles at contact surface make clustering and we can not calculate correctly. Therefore, our method is valid even in the case where the sign of pressure changes spatially and temporary.

\subsection{Equilibrium test with random noise}
In this subsection, we calculate the time evolution from a noisy initial condition in three dimensions. This test is very similar to that of \cite{Dehnen-Aly2012}. Initially, the particles are put on a face-centered cubic lattice with nearest-neighbor distance $d_{{\rm nn}}=0.1\frac{\sqrt{2}}{2}$. We add random noise of the position with the amplitude of $0.1 d_{{\rm nn}}$ to each direction of each particle. Initial velocities of all particles are set to $0$. We put $4000$ particles in the computational domain, and assume a periodic boundary condition. 

We set the mass of each particle $m=2.5 \times 10^{-4}$. Thus the density in unperturbed state is $\rho_{0}=1.0$. We use equation of state Eq.\,(\ref{EoS}), and $C_{s}=1.0$, $\rho_{0,{\rm eos}}=0.8$ for positive pressure, $\rho_{0,{\rm eos}}=1.2$ for negative pressure. We assume constant smoothing length with $h=d_{{\rm nn}}$, and we use Riemann solver for Eq.\,(\ref{EoS}) to achieve the equilibrium state.  

Following the test of \cite{Dehnen-Aly2012}, we calculate the time evolution of $q_{{\rm min}}$, which is defined as,

\begin{equation}
q_{{\rm min}}\equiv \min_{{\rm all \ pairs \ of \ }ij}|\bm{r}_{i}-\bm{r}_{j}|/d_{{\rm nn}}. 
\label{define-qmin}
\end{equation}

\begin{figure}[!htb]
  \begin{center}
    \includegraphics[width=9cm,height=6cm]{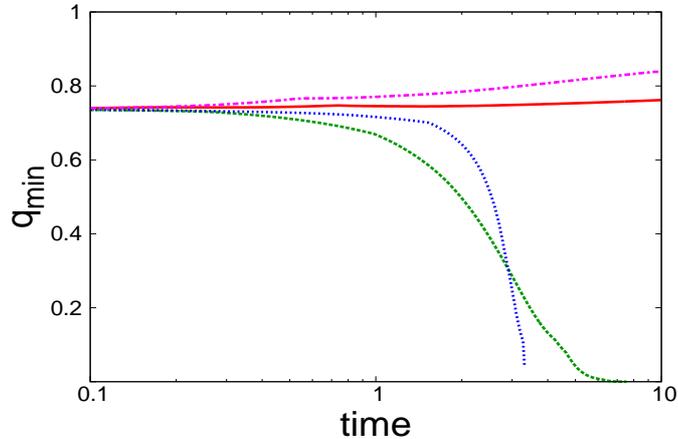}
    \caption{Time evolution of $q_{{\rm min}}$ for the simulations starting from noisy initial conditions in three dimensions. The red solid line shows linear interpolation with positive pressure, the green dashed line shows cubic spline interpolation with positive pressure, the blue dotted line shows linear interpolation with negative pressure, and the pink chain line shows cubic spline interpolation with negative pressure.}
    \label{3D-qmin-test}
  \end{center}
\end{figure}

\noindent This $q_{{\rm min}}$ is an indicator for the regularity of particle distribution. If the particles are put on a perfect face-centered cubic lattice, $q_{{\rm min}}$ becomes one, while if the particles are clustered this value is close to $0$. In a typical grass-like distribution (a uniform-density equilibrium distribution), $q_{{\rm min}}\sim 0.7$. Figure \ref{3D-qmin-test} shows the time evolution of $q_{{\rm min}}$ in the case of linear interpolation and cubic spline interpolation for positive and negative pressure respectively.

As shown in Fig.\,\ref{3D-qmin-test}, $q_{{\rm min}}$ of linear interpolation for negative pressure and cubic spline interpolation for positive pressure become close to $0$ during the time evolution, which means that the particles are clustering. On the other hand, $q_{{\rm min}}$ of linear interpolation for positive pressure remains almost initial value. Moreover, that of cubic spline interpolation for negative pressure increases gradually and becomes close to $1$, which means that the distribution of particles gradually settles onto a face-centered cubic lattice. Therefore, these results in three-dimensional simulation also agree with those of linear stability analysis.

\section{Summary and Future Work}
In the SPH method, there is a numerical instability that results in the clustering of particles under certain conditions, and this tensile instability is significant in the negative pressure regime. In this paper we show that the formalism of the Godunov SPH method \cite{Inutsuka2002} is viable for mitigating the tensile instability. We formulate higher-order approximations for the interpolation of $V(s)$ and conduct linear stability analyses for the various equations of motion of the Godunov SPH method. 

We conclude that in the one-dimensional case the stable interpolations are linear interpolation and cubic spline interpolation for positive pressure media, and quintic spline interpolation for negative pressure media. In the two- and three-dimensional cases, linear interpolation is stable for positive pressure, and cubic spline interpolation and quintic spline interpolation are stable for negative pressure. In the case of variable smoothing length, calculations with sufficiently large $C_{{\rm smooth}}$ remain stable. Therefore, we can suppress the tensile instability by an appropriate order of interpolation without any additional and artificial terms. Additionally, we have derived the analytical solution of the Riemann problem for the equation of state given by Eq.\,(\ref{EoS}), and we confirmed that our analytical solution agrees with the result of the numerical simulation of the shock tube problem.

In practical calculations of elastic dynamics, we need to formulate how to handle deviatoric stress tensor that corresponds to the non-diagonal parts of the stress tensor (e.g., \cite{Monaghan1999,Benz-Asphaug1995}). The extension of the present method to the non-diagonal stress tensor will be studied in our next paper.

\section*{Acknowledgement}
The authors thank Hiroshi Kobayashi, Kazunari Iwasaki, Yusuke Tsukamoto, and Jennifer M. Stone for useful discussions and comments. 

SI is supported by Grant-in-Aid for Scientific Research (23244027, 23103005).

\bibliography{mybibfile}

\section*{Appendix A}
In Appendix A, we explain how to obtain the dispersion relation from the acceleration term. We assume that the wave number vector is along the $x$-direction.

First, we define the unperturbed positions of SPH particles as

\begin{equation}
\overline{\bm{r}_{i}}=(\overline{x_{i}},\overline{y_{i}},\overline{z_{i}}) \equiv (l\Delta x,m\Delta x,n\Delta x), \ \ i=1, \ 2, \cdots N \tag{A1}
\label{A1}
\end{equation}

\noindent where $N$ shows the number of particles, and $l$, $m$, $n$ denote integers 1, 2, $\cdots$. In other words, SPH particles are put on a square lattice with a side length of $\Delta x$ in the unperturbed state. 

We consider the perturbed positions to be

\begin{align}
\bm{r}_{i}=(\overline{x_{i}}+\delta x_{i},\overline{y_{i}},\overline{z_{i}}), \nonumber \\ \delta x_{i}=\epsilon \exp [i(k\overline{x_{i}}-\omega t)], \tag{A2} \label{A2}
\end{align}

\noindent where $k$ is the wave number of the perturbation, $\omega$ is the angular frequency of the perturbation, $\epsilon$ is the amplitude, and $i$ not in subscript shows the imaginary unit. 

We expect that the acceleration can be expressed as

\begin{equation}
\bm{a}_{i}=(a_{x,i},a_{y,i},a_{z,i})=\frac{d^{2}}{dt^{2}}\bm{r}_{i}=(-\omega^{2}\delta x_{i},0,0). \tag{A3}
\label{A3}
\end{equation}

In practice, we can calculate $\omega^{2}$ by taking the ratio of $a_{x,i}$ and $\delta x_{i}$:

\begin{equation}
  \omega^{2}=-a_{x,i}/\delta x_{i}. \tag{A4}
\label{A4}
\end{equation}

Thus we can obtain the dispersion relation by taking the ratio of the displacement and the acceleration for various wave numbers $k$: 

\begin{equation}
\omega^{2}={\rm Re} \Bigl[\frac{1}{N}\sum_{i=1}^{N}\Bigl( -\frac{a_{x,i}}{\delta x_{i}} \Bigr) \Bigr]. \tag{A5}
\label{A5}
\end{equation}

\section*{Appendix B}
In Appendix B, we describe the linear stability analysis of the Godunov SPH method.

We assume particle positions as in Eq.\,(\ref{A2}). Here we consider $\epsilon$ as an infinitesimal constant and neglect second or higher orders of $\epsilon$. Only the $x$-component of the specific-volume gradient and the acceleration appear, and the other components vanish because the perturbation is only along the $x$-axis. We assume infinitely accurate time integration, and ignore the effect of the discretization of time integration.

We assume the masses of all particles, $m$, are the same. We use the equation of motion given by Eq.\,(\ref{EoM-consth}), and use Eq.\,(\ref{P-inviscid}) for $P_{ij}^{\ast}$. For $V_{ij}^{2}$, we use Eq.\,(\ref{Vij2-linear}) for linear interpolation, Eq.\,(\ref{Vij2-cubic}) for cubic spline interpolation, and Eq.\,(\ref{Vij2-quintic}) for quintic spline interpolation.

First, we linearize the density of particle $i$ using Eq.\,(\ref{density}):

\begin{align}
\rho_{i}&=\sum_{j}mW(\bm{r}_{i}-\bm{r}_{j},h) \approx \sum_{j}mW(\overline{\bm{r}_{i}}-\overline{\bm{r}_{j}},h)+\sum_{j}m(\delta x_{i}-\delta x_{j})\frac{\partial}{\partial \overline{x_{i}}}W(\overline{\bm{r}_{i}}-\overline{\bm{r}_{j}},h) \nonumber \\ &=\sum_{j}mW(\overline{\bm{r}_{i}}-\overline{\bm{r}_{j}},h)+\Bigl[\sum_{j}m(1-\exp [-ik(\overline{x_{i}}-\overline{x_{j}})])\frac{\partial}{\partial \overline{x_{i}}}W(\overline{\bm{r}_{i}}-\overline{\bm{r}_{j}},h)\Bigr]\delta x_{i}. \tag{B1} \label{B1}
\end{align}

\noindent The first term on the right-hand side of Eq.\,(\ref{B1}) is the density of the unperturbed state. This term is almost the same as the average density $\rho_{0}$. The terms with odd functions of $\overline{x_{i}}-\overline{x_{j}}$ vanish when we sum over subscript $j$. Thus, Eq.\,(\ref{B1}) becomes

\begin{equation}
\rho_{i}\approx \rho_{0}(1-iD\delta x_{i}), \tag{B2} 
\label{B2}
\end{equation}

\noindent where $D$ is defined by Eq.\,(\ref{DR-linear-cubic-2-3D-coefficients}). From Eq.\,(\ref{B2}), we can immediately find the linearized specific volume as

\begin{equation}
V_{i}=\frac{1}{\rho_{0}}(1+iD\delta x_{i}). \tag{B3}
\label{B3}
\end{equation}

Next, we linearize the $x$-component of the gradient of the specific volume. From Eq.\,(\ref{second-differencial-V}), the gradient of the specific volume becomes

\begin{align}
\frac{\partial V_{i}}{\partial x_{i}}&=-\frac{1}{\rho_{i}^{2}}\frac{\partial \rho_{i}}{\partial x_{i}} \nonumber \\ & \approx -\frac{1}{\rho_{0}^{2}}\sum_{j}m(\delta x_{i}-\delta x_{j})\frac{\partial^{2}}{\partial \overline{x_{i}}^{2}}W(\overline{\bm{r}_{i}}-\overline{\bm{r}_{j}},h) =-\frac{1}{\rho_{0}}C_{\rho}\delta x_{i}, \tag{B4} \label{B4}
\end{align}

\noindent where $C_{\rho}$ is defined by Eq.\,(\ref{DR-linear-cubic-2-3D-coefficients}). Finally, we linearize the second-order derivative of the specific volume using Eq.\,(\ref{second-differencial-V}):

\begin{align}
\frac{\partial^{2}V_{i}}{\partial x_{i}^{2}}=\sum_{j}\frac{m}{\rho_{j}}\frac{\partial V_{j}}{\partial x_{j}}\frac{\partial}{\partial x_{i}}W(\bm{r}_{i}-\bm{r}_{j},h)\approx -\frac{i}{\rho_{0}}C_{\rho}D\delta x_{i} \tag{B5}. \label{B5}
\end{align}

The linearized pressure of particle $i$ is

\begin{equation}
P_{i}=P_{0}+\delta P \approx P_{0}+C_{s}^{2}\delta \rho = P_{0} -iC_{s}^{2}\rho_{0}D\delta x_{i}. \tag{B6}
\label{B6}
\end{equation}

We substitute these linearized physical quantities into Eq.\,(\ref{EoM-consth}), and we define the coefficients $a$, $b$, and so on using Eq.\,(\ref{DR-linear-cubic-2-3D-coefficients}). Finally, we obtain the analytical solution of $\omega^{2}$, such as Eq.\,(\ref{DR-linear-2-3D}), Eq.\,(\ref{DR-cubic-2-3D}), and Eq.\,(\ref{DR-quintic-2-3D}).

\end{document}